\documentclass[aps,prx,reprint,showkeys,longbibliography,superscriptaddress]{revtex4-2}
\usepackage{amsmath,amssymb,amsthm,epsfig,bbm,bm,verbatim,MnSymbol,color}
\usepackage[colorlinks=true,citecolor=blue]{hyperref}

\usepackage{graphicx}
\usepackage{dcolumn}
\usepackage{bm}
\usepackage{hyperref}

\usepackage{physics}
\usepackage{ytableau}

\newcommand{\mc}[1]{\mathcal{#1}}
\newcommand{\id}{\mathbbm{1}}
\renewcommand{\tr}[1]{\Tr[#1]}

\newtheorem{theorem}{Theorem}
\newtheorem{corollary}{Corollary}
\newtheorem{lemma}{Lemma}

\begin{document}

\title{Thermodynamics of Permutation-Invariant Quantum Many-Body Systems:\\ A Group-Theoretical Framework}

\author{Benjamin Yadin}
\affiliation{Naturwissenschaftlich-Technische Fakult\"at, Universit\"at Siegen, Walter-Flex-Straße 3, 57068 Siegen, Germany}

\author{Benjamin Morris}
\affiliation{School of Physics and Astronomy, University of Nottingham, Nottingham NG7 2RD, United Kingdom}
\affiliation{Centre for the Mathematics and Theoretical Physics of Quantum Non-equilibrium Systems, University of Nottingham, Nottingham NG7 2RD, United Kingdom}

\author{Kay Brandner}
\affiliation{School of Physics and Astronomy, University of Nottingham, Nottingham NG7 2RD, United Kingdom}
\affiliation{Centre for the Mathematics and Theoretical Physics of Quantum Non-equilibrium Systems, University of Nottingham, Nottingham NG7 2RD, United Kingdom}

\date{\today}

\begin{abstract}
Quantum systems of indistinguishable particles are commonly described using the formalism of second quantisation, which relies on the assumption that any admissible quantum state must be either symmetric or anti-symmetric under particle permutations. 
Coherence-induced many-body effects such as superradiance, however, can arise even in systems whose constituents are not fundamentally indistinguishable as long as all relevant dynamical observables are permutation-invariant. 
Such systems are not confined to symmetric or anti-symmetric states and therefore require a different theoretical approach.  
Focusing on non-interacting systems, here we combine tools from representation theory and thermodynamically consistent master equations to develop such a framework. 
We characterise the structure and properties of the steady states emerging in permutation-invariant ensembles of arbitrary multi-level systems that are collectively weakly coupled to a thermal environment.
As an application of our general theory, we further explore how these states can in principle be used to enhance the performance of quantum thermal machines. 
Our group-theoretical framework thereby makes it possible to analyse various limiting cases that would not be accessible otherwise. 
In addition, it allows us to show that the properties of multi-level ensembles differ qualitatively from those of spin ensembles, which have been investigated earlier using the standard Clebsch-Gordan theory. 
Our results have a large scope for future generalisations and pave the way for systematic investigations of collective effects arising from permutation-invariance in quantum thermodynamics. 
\end{abstract}

\maketitle

\section{Introduction}
    
    Permutation symmetry plays a fundamental role in many-body quantum mechanics. 
    According to the principle of \emph{indistinguishability}, quantum states that differ only by interchanges of identical particles cannot be distinguished by any measurement \cite{Massiah1964,Ballentine1998}.
    As a result, any dynamical observable of a system of identical particles, including its Hamiltonian, must be invariant under particle permutations. 
    The eigenstates of such observables can be divided into linearly independent sets according to their symmetry type under permutations -- for two particles, the eigenstates of a permutation-invariant observable can be chosen as either symmetric or anti-symmetric; for more than two particles, additional, partially symmetric types arise. 
    These symmetry types are preserved under the time evolution of the system. 
    Moreover, they give rise to super-selection rules which forbid transitions between states of different types and imply that coherences between such states cannot be observed. 
    
    \begin{figure}[h!]
        \includegraphics[width=.47\textwidth]{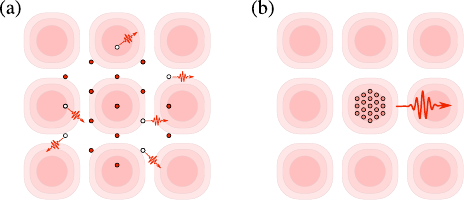}
        \caption{\label{fig:Intro}
        Permutation-invariant ensembles. A collection of two-level atoms, which can be externally distinguished by their position in a lattice, is coupled to a thermal radiation field whose spatial intensity is represented by the red shaded areas. 
        The system is initially prepared in a fully symmetric state, where all atoms are excited. 
        (a) If the lattice spacing is larger than the typical wave length of the radiation field, the atoms decay independently under the spontaneous emission of photons. 
        At long times, the ensemble settles to a Gibbs state that is a statistical mixture of many-body quantum states with all possible permutation symmetry types.
        (b) When its typical wave length is comparable to the lattice spacing, the atoms become indistinguishable to the radiation field. As a result, the permutation symmetry of the initial state must be preserved during the decay process, which leads to the collective emission of radiation, i.e., superradiance~\cite{Dicke1954}. The ensemble thereby approaches a non-thermal steady state, which contains only fully symmetric quantum states.}
    \end{figure}

    The \emph{symmetrisation postulate} is more restrictive than the principle of indistinguishability. 
    It states that, depending on the sort of particle, only quantum states with one specific symmetry type exist: identical Bosons are described by symmetric states, identical Fermions by anti-symmetric ones \cite{Ballentine1998}. 
    This restriction has profound consequences. 
    It implies, for example, that the particles of an ideal quantum gas must be distributed in energy space according to either Bose-Einstein or Fermi-Dirac statistics, which lead to fundamentally different physical properties.  
    On a mathematical level, the symmetrisation postulate is elegantly accommodated by the formalism of second quantisation, which inherently restricts the Hilbert space of a many-body system to its fully symmetric or anti-symmetric subspace.

    Here we focus on many-body systems that are not subject to the symmetrisation postulate but whose relevant dynamical observables are still permutation-invariant. 
    Specifically, we consider collections of non-interacting multi-level systems which are coupled to an environment that cannot distinguish between their constituents. 
    The particles may be not fundamentally identical as an observer may still be able to tell them apart through an external degree of freedom, see Fig.~\ref{fig:Intro}.
    For clarity, we will refer to this type of many-body system as \emph{permutation-invariant ensembles}. 
    Such settings can be realised for instance with cold atoms \cite{Norcia2016-lt,Norcia2018-sx}, ions \cite{Casabone2015-ex} or artificial atoms \cite{Scheibner2007-ss} and can host remarkable phenomena with \emph{superradiance} being a prime example \cite{Dicke1954}.

    Such \emph{collective effects} in many-body systems have recently received much attention in quantum thermodynamics, both in the context of fundamental questions \cite{Bengtsson2018,Plesch2015Maxwell,Braun2018Quantum,Yadin2021Mixing} and on the more applied side, for instance as a means of enhancing the performance of quantum heat engines and refrigerators \cite{Jaramillo_2016,Macovei2022,Uzdin2016,Myers2020,Myers_2022,Latune2019-oi,Mukherjee_2021,Niedenzu_2018,Latune2019,Latune2020-ex,Souza2021,Kloc2021,Kaimimura2022,Mayo2022,Kim2022Photonic} or isothermal machines \cite{Li_2018,Chen_2019,Keller2020,Fogarty_2020,Carollo2020a,Carollo2020b,Myers2021,Watanabe2020Quantum}.
    Studies of devices featuring strong inter-particle interactions  \cite{Jaramillo_2016,Macovei2022,Li_2018,Chen_2019,Keller2020,Fogarty_2020,Carollo2020a,Carollo2020b,Myers2021} are usually based on specific models, since interacting many-body systems are notoriously hard to describe on a general level, especially when coupled to a thermal environment. 
    Indistinguishability-induced collective thermodynamical effects can be observed even in non-interacting systems
    \cite{Uzdin2016,Myers2020,Myers_2022,Latune2019-oi,Mukherjee_2021,Niedenzu_2018,Latune2019,Latune2020-ex,Souza2021,Kloc2021,Kaimimura2022,Mayo2022},
    which in many cases admit a comparatively simple theoretical description in terms of thermodynamically consistent quantum master equations \cite{Kosloff2013,Brandner2016}. 
    This approach, however, has so far been mostly limited to spin ensembles \cite{Niedenzu_2018,Latune2019,Latune2020-ex,Souza2021,Kloc2021,Kaimimura2022,Mayo2022}, whose collective behaviour is captured by standard Clebsch-Gordan theory. 
    In developing a systematic generalisation of this theory to arbitrary multi-level systems, our work provides a universal framework to further investigate the principal role of collective effects in quantum thermodynamical processes and to uncover their potential applications. 
    
    The quantum states of a permutation-invariant ensemble are generally not restricted to a specific symmetry type and span the entire many-body Hilbert space, which renders the formalism of second quantisation inapplicable. 
    Our first major aim is to show that such systems nevertheless admit an efficient theoretical description, which, among other results, provides explicit means to calculate the expectation values of permutation-invariant observables. 
    At the heart of this framework lies \emph{Schur-Weyl duality}, a powerful tool from representation theory, which makes it possible to endow the Hilbert space of a permutation-invariant ensemble with a universal structure \cite{Goodman2009,Rowe2010}.   
    Specifically, this structure consists of a series of invariant subspaces that are associated with both the permutation group and the symmetry group generated by the dynamical variables of the system. 
    For spin systems, these dynamical variables correspond to angular momentum degrees of freedom, which generate the quantum rotation group $\mathrm{SU}(2)$. 
    In this special case, which has been studied before \cite{Niedenzu_2018,Latune2019,Latune2020-ex,Souza2021,Kloc2021,Kaimimura2022,Mayo2022} and will serve as a reference for our analysis, the Schur-Weyl decomposition reduces to the conventional Clebsch-Gordan series \cite{Ballentine1998}. 
    
    As a consequence of the super-selection rules implied by the indistinguishability principle, permutation-invariant ensembles do generally not relax to a Gibbs state in a thermal environment. 
    Instead, they settle to a non-trivial steady state that still depends on the initial state in which the ensemble was originally prepared, see Fig.~\ref{fig:Intro}. 
    Our second major aim is to systematically characterise these steady states, determine their thermodynamical properties, analyse their internal structure and explore how they can be utilised to enhance the performance of quantum engine cycles.  
    Our mathematical framework does however have a wider scope, which extends beyond steady states.
    It thus opens an interesting perspective for future research.

A second potential area of application for our mathematical framework is quantum information theory, where permutation-invariant systems have been studied in the context of entanglement and reference frames~\cite{Jones2006Entanglement}, as well as for describing hidden degrees of freedom in quantum optics~\cite{Adamson2008Detecting,Turner2016Postselective,Stanisic2018Discriminating}.
Similarly, bunching effects in quantum optics have been shown to have implications for energy transfer between light and a mechanical oscillator~\cite{Holmes2020Enhanced,Holmes2020Gibbs}.
In general, there is potential for exploring a variety of quantum settings in which the same permutation-invariant structures arise.

The paper is structured as follows.
In Section~\ref{sec:MathThools}, we outline the mathematical background necessary for the analysis.
Readers familiar with representation theory may skip this section.
In Section~\ref{sec:setup}, we describe the physical setting and weak-coupling model of thermalisation in open systems, giving the general partially thermalised form of the steady state.
We give computable formulas for basic thermodynamical quantities in Section~\ref{sec:ThermoQuant} and then put these to use in Section~\ref{sec:engine} by studying a model of a heat engine, the Otto cycle.
Finally, in Section~\ref{sec:higher-order}, we explore the non-classical implications of higher-order symmetry groups beyond spins, followed by broader perspectives in Section~\ref{sec:perspectives}.

\section{Mathematical Tools}\label{sec:MathThools}

In this section, we introduce the mathematical tools that will be used in our analysis. We provide a brief guide to the Lie group $\mathrm{SU}(d)$, its Lie algebra $su(d)$ and their representations. We further discuss Schur-Weyl duality, Young diagrams, group characters and Weyl's character formula. To illustrate these general concepts, we show how they appear in the common theory of angular momentum. Readers already familiar with these mathematical tools may skip to the next section.

\subsection{The special unitary group}

The special unitary group of degree $d$, denoted by $\mathrm{SU}(d)$, is the group of unitary $d\times d$-matrices with determinant $1$. 
Any element $u$ of this group can be written in the form $u=e^{i h}$, where $h$ is traceless Hermitian $d\times d$-matrix. 
These matrices form a vector space over the real numbers. 
Hence, upon choosing a basis $\{x_a\}$ in this space, we can express any $u\in \mathrm{SU}(d)$ as $u= \exp[{i\sum_a \theta_a x_a}]$ with $a=1,\dots, d^2-1$ and $\theta_a\in\mathbb{R}$. 
The matrices $x_a$ are called \emph{generators} of $\mathrm{SU}(d)$. 
They satisfy a set of characteristic commutation relations
\begin{equation}\label{equ:cancomm}
    [x_a,x_b] = i\sum_c f^{ab}_c x_c, 
\end{equation}
which define the Lie algebra $su(d)$ with the coefficients $f^{ab}_c = - f^{ba}_c \in \mathbb{R}$ being called \emph{structure constants} (taking values depending on the chosen basis).  

In practice, it is often convenient to chose a special non-Hermitian basis for the Lie algebra $su(d)$, which is known as the \emph{Cartan basis} \cite{Zee2016}. 
To construct this basis, we first define a set of $d-1$ diagonal Hermitian generators $d_i$, which satisfy 
\begin{align}\label{equ:comm1}
[d_i,d_j]=0. 
\end{align}
These generators form the Cartan subalgebra of $su(d)$. 
The remaining $d(d-1)$ generators, which we denote by $e_\mu$, can be chosen such that they obey the commutation relations 
\begin{align}\label{equ:comm2}
    [d_i,e_\mu]=v^i_\mu e_\mu, 
\end{align}
where the constants $v_\mu^i \in \mathbb{R}$ form the \emph{root vector} $\mathbf{v}_\mu =(v_\mu^1,\dots, v_\mu^{d-1})$ associated with the non-diagonal generator $e_\mu$.
For simplicity we will often refer to $\mu$ as a root.

The non-diagonal generators $e_\mu$ come in Hermitian conjugate pairs. 
As can be seen from Eq.~\eqref{equ:comm2}, they obey the symmetry $e_\mu^\dagger = e_{-\mu}$ with $\mathbf{v}_{-\mu} = -\mathbf{v}_\mu$. 
In the Cartan basis, the exponential map that connects the group $\mathrm{SU}(d)$ with its Lie algebra $su(d)$ reads $u = \exp [ i (\sum_i \alpha_i d_i + \sum_\mu \gamma_\mu e_\mu)]$, where $\alpha_i\in\mathbb{R}$ and $\gamma_\mu^\ast = \gamma_{-\mu}\in\mathbb{C}$. 
Using the chain rule for the commutator, one can now derive the remaining commutation relations 
\begin{align}
    &[e_\mu,e_\nu]=N_{\mu\nu}\,e_{\mu+\nu}\quad\text{for}\quad \nu\neq -\mu \quad\text{and}\label{equ:comm3a}\\ 
    &[e_\mu,e_{-\mu}]= \sum_i M_{\mu i}\, d_i\label{equ:comm3b}, 
\end{align}
which, together with the relations \eqref{equ:comm1} and \eqref{equ:comm2}, fully specify the Lie algebra $su(d)$ in the Cartan basis. 
Here, $\mu+\nu$ denotes a root with corresponding root vector $\mathbf{v}_{\mu+\nu} := \mathbf{v}_\mu + \mathbf{v}_\nu$.
The coefficients $N_{\mu\nu}, M_{\mu i}\in\mathbb{R}$ depend on the normalization of the generators, where the $N_{\mu\nu}$ obey the symmetry $N_{\mu\nu}=-N_{\nu\mu}$ and vanish if there exists no generator with the root vector $\mathbf{v}_{\mu+\nu}$.

The $d(d-1)$ root vectors $\mathbf{v}_\mu$ are in one-to-one correspondence with the generators $e_\mu$ and span a $(d-1)$-dimensional vector space and therefore cannot all be linearly independent. 
In fact, one can always find a minimal generating set of roots $\hat{\mathbf{v}}_\mu$ such that any root vector can be decomposed as $\mathbf{v}_\mu = \sum_{\nu=1}^{d-1} n_{\mu\nu}\hat{\mathbf{v}}_\nu$, where the coefficients $n_{\mu\nu}$ are integers, which are either all non-positive or non-negative. 
The vectors $\hat{\mathbf{v}}_{\mu}$ are then referred to as \emph{simple roots} \cite{Zee2016}. 
A complete set of simple roots has $d-1$ elements (and there is in general no unique choice).

In physics, the Hermitian generators of $\mathrm{SU}(d)$ are related to the dynamical variables of a $d$-level system. 
For example, a Hermitian basis of $su(2)$ is given by
\begin{equation}\label{equ:Pauli}
    s_x = \frac{1}{2} \begin{bmatrix}
         0 & 1\\ 1 & 0
    \end{bmatrix},\quad
    s_y = \frac{1}{2} \begin{bmatrix}
         0 & -i\\ i & 0
    \end{bmatrix},\quad 
    s_z = \frac{1}{2} \begin{bmatrix}
         1 & 0\\ 0 & -1
    \end{bmatrix}.
\end{equation}
These matrices represent, up to a factor $\hbar$, the dynamical variables of a spin-$\frac{1}{2}$ system. 
Upon setting $x_1 = s_x$, $ x_2=s_y$ and $x_3=s_z$, the generators satisfy the commutation relations \eqref{equ:cancomm} with the structure constants $f^{ab}_c = \varepsilon_{abc}$, where $\varepsilon_{abc}$ denotes the Levi-Civita symbol. 
The Cartan basis of $su(2)$ is usually constructed by choosing the diagonal generator as $d_1 = s_z$. 
The non-diagonal generators are then given by the raising and lowering operators $e_\pm = s_x \pm i s_y$, which obey the commutation relations \eqref{equ:comm2} with the one-dimensional root vectors $\mathbf{v}_\pm = \pm 1$;
the coefficients appearing in Eq.~\eqref{equ:comm3b} are given by $M_{\pm 1}= \pm 1/2$.
Since $\mathbf{v}_- = - \mathbf{v}_+$, either $\mathbf{v}_+$ or $\mathbf{v}_-$ can be chosen as a simple root of $su(2)$. 

As a second example, we consider $su(3)$, whose two diagonal generators can be identified with the Gell-Mann matrices 
\begin{equation}\label{equ:GellMann}
\Lambda_3 = \left[\begin{array}{lll}
				1 & 0 & 0\\
				0 & -1 & 0\\
				0 & 0 & 0
			\end{array}\right], \quad
\Lambda_8 = \frac{1}{\sqrt{3}} \left[\begin{array}{lll}
				1 & 0 & 0\\
				0 & 1 & 0\\
				0 & 0 & -2
			\end{array}\right], 
\end{equation}
i.e., $d_1=\Lambda_3$ and $d_2=\Lambda_8$. 
In relativistic particle physics, these generators are associated with the dynamical variables \textit{Isospin} and \textit{Hypercharge}. 
A complete set of simple roots for $su(3)$ is given by $\hat{\mathbf{v}}_{+1} = (2,0),\, \hat{\mathbf{v}}_{+2} = (1, \sqrt{3})$ and the corresponding non-diagonal generators have the matrix form
\begin{equation}\label{equ:su3_simple_roots}
    e_{+1} = \left[\begin{array}{lll}
                    0 & 1 & 0\\
                    0 & 0 & 0\\
                    0 & 0 & 0
                \end{array}\right], \quad
    e_{+2} =  \left[\begin{array}{lll}
                    0 & 0 & 1\\
                    0 & 0 & 0\\
                    0 & 0 & 0
                \end{array}\right]. 
\end{equation}

\subsection{Irreducible representations }

A unitary representation of $\mathrm{SU}(d)$ on some Hilbert space $\mathcal{H}$ associates every $u\in \mathrm{SU}(d)$ with a unitary operator $U(u)$ such that $U(u)U(u')= U(uu')$. 
If $\mathcal{H}$ has no non-trivial subspace that is invariant under the action of all operators $U(u)$, the representation is called \emph{irreducible}. 
Any reducible, i.e.\ not irreducible, representation can be decomposed into a direct sum of irreducible representations, or irreps.
That is, there exists an orthonormal basis of $\mathcal{H}$, in which $U(u)$ takes the form
\begin{equation}\label{equ:redecomp}
    U(u) = \bigoplus_\lambda U^\lambda (u)
\end{equation}
with the operators $U^\lambda (u)$ forming irreps of $\mathrm{SU}(d)$ on some orthogonal subspaces $\mathcal{H}^\lambda$ of $\mathcal{H}$.
The label $\lambda$ thereby represents an ordered partition of some integer $n$ into $d$ non-negative integers $\lambda_1\geq \lambda_2 \geq \cdots\geq \lambda_d$. 
The decomposition \eqref{equ:redecomp} may contain multiple terms with the same label $\lambda$. 
We note that the natural representation $U(u)=u$ of $\mathrm{SU}(d)$ on $\mathcal{H}=\mathbb{C}^d$ is always irreducible. 

The concept of representations naturally extends to the generators: any $U(u)$ can be expressed as $U(u)=\exp[i\sum_a \theta_a X_a]$, where $\theta_a\in\mathbb{C}$ and the $X_a$ are traceless operators satisfying the same commutation relations as the generators $x_a$, i.e. 
\begin{align}
[X_a,X_b]=i \sum_c f^{ab}_c X_c
\end{align}
with the same structure constants $f^{ab}_c$ as in Eq.~\eqref{equ:cancomm}. 
Hence, the operators $X_a$ form a representation of the Lie algebra $su(d)$ on $\mathcal{H}$. 
These operators can be chosen to be Hermitian, in which case the coefficients $\theta_a$ must be real. 
In general, a representation can be constructed for any basis of $su(d)$. 
In particular, we may pick the Cartan basis, whose representation consists of $d-1$ Hermitian operators $D_i$ and  $d(d-1)$ operators $E_\mu$ that come in Hermitian conjugate pairs satisfying $E_\mu^\dagger = E_{-\mu}$. 
Since these operators obey the same commutation relations as the corresponding generators $d_i$ and $e_\mu$, cf.  Eqs.~\eqref{equ:comm1}-\eqref{equ:comm3b}, there exists a basis of $\mathcal{H}$, in which the $D_i$ become diagonal and the the $E_\mu$ assume the roles of raising and lowering operators. 

Any irrep of the group $\mathrm{SU}(d)$ must derive from an irrep of the Lie algebra $su(d)$. 
That is, if the operators $U(u)$ do not share a non-trivial invariant subspace, the same must be true for the corresponding representations $X_a$ of the generators. 
The converse statement also holds: any irrep $X_a$ of $su(d)$ gives rise to an irrep of $\mathrm{SU}(d)$ via the exponential map $U(u)= \exp[i\sum_a \theta_a X_a]$. 
Any reducible representation of $\mathrm{SU}(d)$ must therefore be associated with a reducible representation of $su(d)$, which can be decomposed into a direct sum of irreps.   
Specifically, in the basis of $\mathcal{H}$ where Eq.~\eqref{equ:redecomp} holds, we also have
\begin{equation}\label{equ:LAblock}
    X_a = \bigoplus_\lambda X^\lambda_a
\end{equation}
with the $X^\lambda_a$ forming an irrep of $su(d)$ on the subspace $\mathcal{H}^\lambda$. 
Since changing the basis of $su(d)$ is a linear operation on both the generators and their representations, Eq.~\eqref{equ:LAblock} holds for any basis of $su(d)$. 

In the previous section, we have seen that the Hermitian generators of $\mathrm{SU}(d)$ can be identified with the fundamental dynamical variables of a physical system. 
This interpretation extends also to higher-dimensional representations of these generators. 
In particular, a Hermitian irrep of $su(2)$ on $\mathbb{C}^{2s+1}$ describes a spin-$s$ system. 
For example, the $3$-dimensional analogues of the matrices \eqref{equ:Pauli}, 
\begin{equation}\label{equ:s1Pauli}
    S_x^1 = \frac{1}{\sqrt{2}}
    \begin{bmatrix}
         0 & 1 & 0\\
         1 & 0 & 1\\
         0 & 1 & 0
    \end{bmatrix}, \;
    S_y^1 = \frac{1}{\sqrt{2}}
    \begin{bmatrix}
         0 & -i & 0\\
         i & 0  & -i\\
         0 & i  & 0
    \end{bmatrix},\;
    S_z^1 =
    \begin{bmatrix}
         1 & 0 & 0\\
         0 & 0 & 0\\
         0 & 0 & -1
    \end{bmatrix}
\end{equation}
form an irrep of $su(2)$ on $\mathbb{C}^3$ with $X_1=S_x^1$, $X_2=S_y^1$ and $X_3=S_z^1$ and represent the dynamical variables of a spin-1 system. 

\subsection{Schur-Weyl duality}\label{sec:SW}

We now consider the representations of $\mathrm{SU}(d)$ on the product space $\mathcal{H}^{(n)}=(\mathbb{C}^d)^{\otimes n}$, which is the Hilbert space of an ensemble of $n$ $d$-level systems. 
We will refer to the individual systems as \emph{particles} in the following.
Extending the natural representation of $\mathrm{SU}(d)$ to $\mathcal{H}^{(n)}$ yields the tensor-product representation $U^{(n)}(u)=u^{\otimes n}$, which is generally reducible. 
Hence, the operators $U^{(n)}(u)$ can be brought to the block-diagonal form of Eq.~\eqref{equ:redecomp}. 
A block with dimension $d_\lambda = \dim \mathcal{H}^\lambda$ can thereby appear multiple times and we denote its multiplicity by $m_\lambda$. 
The basis of $\mathcal{H}^{(n)}$ that produces the decomposition \eqref{equ:redecomp} is called the \emph{Schur basis}. 
Finding this basis is generally complicated. 
However, the dimensions $d_\lambda$ and multiplicities $m_\lambda$ of the individual irreps can be calculated without explicit knowledge of the Schur basis, as we will see in the next section.
In this section, we first introduce the mathematical tool that makes these calculations possible. 

\emph{Schur-Weyl duality} asserts that the tensor-product representation $U^{(n)}(u)$ of $\mathrm{SU}(d)$ admits the decomposition 
\begin{equation}\label{equ:SW}
    U^{(n)}(u) = \bigoplus_\lambda \mathbbm{1}_{\mathcal{K}^\lambda}\otimes U^\lambda (u),
\end{equation}
where $\lambda$ stands for an ordered partition of $n$ into $d$ integers, $U^\lambda (u)$ is an irrep of $\mathrm{SU}(d)$ with dimension $d_\lambda$ and $\mathbbm{1}_{\mathcal{K}^\lambda}$ is the identity operator on an $m_\lambda$-dimensional Hilbert space $\mathcal{K}^\lambda$~\cite{Goodman2009,Rowe2010}. 
As will be explained in Sec.~\ref{sec:young_diagrams}, each $\lambda$ indicates a different symmetry type of the wavefunction.

More generally, Schur-Weyl duality states that the product space $\mathcal{H}^{(n)}$ decomposes as
\begin{equation}
\mathcal{H}^{(n)}=\bigoplus_\lambda \mathcal{K}^\lambda\otimes \mathcal{H}^\lambda, 
\end{equation}
where $\mathcal{K}^\lambda$ and  $\mathcal{H}^\lambda$ carry irreps of the permutation group over $n$ elements $S_n$ and the unitary group $U(d)$ respectively \cite{Goodman2009,Rowe2010}. 
In the following, we will therefore refer to $\mathcal{K}^\lambda$ as a \emph{permutation subspace} and to $\mathcal{H}^\lambda$ as a \emph{unitary subspace}. 
In practical terms, Schur-Weyl duality implies that any operator $O$ on the product space $\mathcal{H}^{(n)}$ that is invariant under arbitrary particle permutations takes the form 
\begin{equation}
O = \bigoplus_\lambda \mathbbm{1}_{\mathcal{K}^\lambda}\otimes O^\lambda
\end{equation}
in the Schur basis. 
The identity \eqref{equ:SW} then follows by noting that the operators $U^{(n)}(u)$ are permutation-invariant and that every irrep of $U(d)$ remains irreducible when being restricted to $\mathrm{SU}(d)$.

We now consider the Lie algebra $su(d)$. 
From a given basis $x_a$, we can construct the tensor-product representation $X^{(n)}_a = \sum_{k=0}^{n-1} \mathbbm{1}_d^{\otimes k}\otimes x_a \otimes\mathbbm{1}_d^{\otimes (n-1-k)}$ of $su(d)$ on $\mathcal{H}^{(n)}$, where $\mathbbm{1}_d$ is the identity matrix of dimension $d$. 
Since operators $X_a^{(n)}$, which generate the tensor-product representation of $\mathrm{SU}(d)$, are permutation-invariant, they decompose in the same way as the operators $U^{(n)}(u)$. 
Hence, in the Schur basis, we have 
\begin{equation}\label{equ:SWGen}
    X_a^{(n)} = \bigoplus_\lambda \mathbbm{1}_{\mathcal{K}^\lambda} \otimes  X_a^{\lambda},
\end{equation}
where the $X_a^{\lambda}$ form an irrep of $su(d)$ with dimension $d_\lambda$. 

Before moving on, it is again instructive to consider the special case of $\mathrm{SU}(2)$. 
The space $\mathcal{H}^{(n)}$ is then the joint Hilbert space of $n$ spin-$\frac{1}{2}$ particles. 
The operators $X^{(n)}_a = S^{(n)}_{x,y,z}=\sum_{k=0}^{n-1} \mathbbm{1}_2^{\otimes k}\otimes s_{x,y,z}\otimes\mathbbm{1}^{\otimes (n-1-k)}_2$, which form a representation of the canonical basis of $su(2)$, correspond to the collective angular momentum operators of the system.
Every ordered partition $\lambda = (\lambda_1,\lambda_2)$ of $n$ can now be uniquely associated with an eigenvalue $[J(J+1)]^\frac{1}{2}$ of the total angular momentum operator $S^{(n)}= \bigl[\mathbf{S}^{(n)}\cdot\mathbf{S}^{(n)}\bigr]^\frac{1}{2}$, where $\mathbf{S}^{(n)}=\bigl(S_x^{(n)},S_y^{(n)},S_z^{(n)}\bigr)$, via the rule $J=(\lambda_1-\lambda_2)/2$. 
Hence, the decomposition \eqref{equ:SWGen} becomes the usual Clebsch-Gordan series, where every irrep of $su(2)$ corresponds to a different value of $J$.

The Schur basis is given by the collective angular momentum basis $\{\ket{J,m,p}\}$, where $m$ is an eigenvalue of $S^{(n)}_z$ and $p$ is an index for the permutation subspace $\mc{K}^J$.
From the theory of angular momenta, we know that $m$ takes the values $-J,-J+1,\dots,J$ for fixed $J$ \cite{Ballentine1998}. 
The irrep $J$ thus has dimension $d_J =2J+1$.
The multiplicities can be determined recursively from the usual rules for the addition of angular momenta.
For $n=1$, we trivially have $J=\frac{1}{2}$ and $m_\frac{1}{2}=1$. 
For two spins, $J$ takes the values 0 and 1 corresponding to singlet and triplet states; hence, we have $m_0=m_1=1$. 
Adding a third spin turns the value 0 into $\frac{1}{2}$, while the value 1 branches into $\frac{3}{2}$ and $\frac{1}{2}$, which gives $m_{\frac{3}{2}}=1$ and $m_{\frac{1}{2}}=2$.  
Analogously, we find $m_0=2$, $m_1=3$ and $m_2=1$ for $n=4$. 
This scheme can, in principle, be applied arbitrarily often, although it becomes cumbersome for large $n$.
In the next section, we describe a more efficient method to calculate the multiplicities of individual irreps for $\mathrm{SU}(2)$ and $\mathrm{SU}(d)$ in general. 

\subsection{Young diagrams} \label{sec:young_diagrams}

An ordered partition $\lambda=(\lambda_1,\dots,\lambda_d)$ of $n$ can be graphically represented as a set of $d$ left-justified rows made up of $\lambda_1,\dots,\lambda_d$ boxes. 
This representation is called the Young diagram $Y_\lambda$ of $\lambda$. 
For instance, the partition $\lambda=(3,1)$ has the Young diagram
\begin{align}\label{equ:YD1}
Y_\lambda=\begin{ytableau}
\,&\,&\,\\ 
\,
\end{ytableau}\; .
\end{align}

The boxes of a Young diagram can be considered as containers for indices that label either single-particle states or particles  \cite{Goodman2009}. 
In the former case, a so-called semi-standard Young tableau is obtained by filling the numbers $1,\dots,d$ into the Young diagram $Y_\lambda$ such that the sequence of numbers in every row is non-decreasing and the sequence in every column is strictly increasing. 
The Young diagram \eqref{equ:YD1} thus admits 3 semi-standard Young tableaux for $d=2$:
\begin{equation}
\begin{ytableau}
1 & 1& 1\\ 
2
\end{ytableau}\;, \;\;\;
\begin{ytableau}
1 & 1& 2\\ 
2
\end{ytableau}\;, \;\;\;
\begin{ytableau}
1 & 2& 2\\ 
2
\end{ytableau}\;. 
\end{equation}
Each Young tableau that is constructed in this way represents a many-particle state with a specific permutation symmetry pattern, which depends only on $\lambda$. 
All of these states are linearly independent and states with different symmetry patterns, i.e.,  states belonging to different Young diagrams, are orthogonal to each other. 
Since the symmetry patterns are invariant under the action of the tensor-product operators $U^{(n)}(u)$ \cite{Goodman2009,Itzykson1966}, the many-particle states derived from a given Young diagram $Y_\lambda$ span an invariant subspace of the $U^{(n)}(u)$. 
Moreover, it can be shown that any invariant subspace of the $U^{(n)}(u)$ can be constructed in this way. 
It follows that the dimensions of these subspaces equal the dimensions $d_\lambda$ of the corresponding irreps $U^\lambda$. 
Thus, $d_\lambda$ can be determined by counting the admissible Young tableaux for a given Young diagram. 
This argument leads to the general formula \cite{Rowe2010}
\begin{align}
    &d_\lambda = \frac{1}{(d-1)!(d-2)!\cdots 1!} \prod_{1\leq i < j\leq d} \left(\tilde{\lambda}_i-\tilde{\lambda}_j\right)
    \quad\text{with}\nonumber\\
    &\tilde{\lambda}= \lambda + \delta \quad\text{and}\quad \delta = (d-1,d-2,\dots,0). 
    \label{equ:dimUlambda}
\end{align}
For $d=2$, we have $d_\lambda = \lambda_1- \lambda_2 +1$ and, upon identifying $J=(\lambda_1-\lambda_2)/2$, we recover the result $d_J = 2J+1$. 

To determine the multiplicities of the irreps $U^\lambda_n$, we have to construct a different set of standard Young tableaux by filling the Young diagrams with the numbers $1,\dots,n$, which label the particles of the ensemble, such that the sequence of numbers in every row and column is strictly increasing. 
For the Young diagram \eqref{equ:YD1}, this rule yields 3 possible standard Young tableaux given by
\begin{equation}
\begin{ytableau}
1 & 2& 3\\ 
4
\end{ytableau}\;, \;\;\;
\begin{ytableau}
1 & 2& 4\\ 
3
\end{ytableau}\;, \;\;\;
\begin{ytableau}
1 & 3& 4\\ 
2
\end{ytableau}\;. 
\end{equation}
The number of possible Young tableaux that can be obtained in this way from a given Young diagram $Y_\lambda$ gives the multiplicity of the corresponding irrep $U^\lambda_n$ and can be determined by combinatorial arguments, which yield the closed-form expression~\cite{Rowe2010}
\begin{equation}\label{equ:multformSW}
    m_\lambda = \frac{n!}{\tilde{\lambda}_1!\cdots\tilde{\lambda}_d!}
    \prod_{1\leq i < j\leq d} \left(\tilde{\lambda}_i-\tilde{\lambda}_j\right).
\end{equation} 
For the special case $d=2$, we thus have, upon substituting $\lambda_1=n/2+J$ and $\lambda_2=n/2-J$, 
\begin{equation}\label{equ:multformSWSpin}
    m_J = \frac{n!(2J+1)}{(n/2+J)!(n/2-J)!}.
\end{equation}
It is easy to check that this formula reproduces the multiplicities that we found iteratively for $n\leq 4$ in the previous section. 

\subsection{Characters}\label{sec:characters}
The final tool that will become essential to our analysis is the character associated to an irrep of $\mathrm{SU}(d)$ \cite{isaacs1994character}. 
In general, the character $\chi_U(u)$ is a function on the group that associates to each element $u$ the trace of the corresponding representation, that is, 
\begin{equation}
    \chi_U(u) = \tr{U(u)}. 
\end{equation}
Here, we are specifically interested in the characters of the Cartan subgroup of $\mathrm{SU}(d)$, whose elements $u_c$ can be expressed solely in terms of the diagonal generators $d_i$. 
That is, we want to calculate objects of the form  
\begin{equation}\label{equ:charcsg}
    \chi_{U^\lambda}(u_c) = \tr{e^{i\sum_i \alpha_i D^\lambda_i}},
\end{equation}
for an irrep $U^\lambda(u)$  of $\mathrm{SU}(d)$ corresponding to the ordered partition $\lambda$ of $n$. 

A general expression for these characters is provided by Weyl's character formula \cite{hall2003lie}, which does not require explicit knowledge of the representation and can be applied as follows. 
We first define the $d\times d$-matices $c_i$, whose $i$th diagonal element is $1$ while all other elements are $0$ \footnote{The matrices $c_i$ form a basis for the diagonal generators of the unitary group $U(d)$.}. 
The diagonal generators $d_i$ can now be expanded in these matrices as $d_i = \sum_j q_{ij} c_j$ with $\sum_j q_{ij} =0$, since the $d_i$ are traceless. 
Once the coefficients $q_{ij}$ have been determined, we can formally express the representations of the diagonal generators as $D^\lambda_i = \sum_j q_{ij} C^\lambda_j$ such that Eq.~\eqref{equ:charcsg} becomes 
\begin{equation}
    \chi_{U^\lambda}(u_c) = \tr{e^{i\sum_j \tilde{\alpha}_j C^\lambda_j}},
\end{equation}
with $\tilde{\alpha}_j=\sum_i \alpha_i q_{ij}$. 
This quantity can now be calculated by means of the formula \cite{greiner2012quantum}
\begin{equation}\label{equ:charform}
    \chi_{U^\lambda}(u_c) = \frac{\det \mathbb{A}[\lambda + \delta]}{\det \mathbb{A}[\delta]},
\end{equation}
where $\delta$ was introduced in Eq.~\eqref{equ:dimUlambda} and $\mathbb{A}$ denotes a $d\times d$-matrix with elements 
\begin{equation}
    \bigl(\mathbb{A}[(r_1,\dots,r_d)]\bigr)_{kl} = e^{i\tilde{\alpha}_k r_l}. 
\end{equation}

To show how this recipe works, we consider again the special case of $\mathrm{SU}(2)$. 
With $d_1=s_z$ the coefficients $q_{ij}$ become $q_{11}=1/2$ and $q_{12}=-1/2$. 
Hence, we have $\tilde{\alpha}_1 = \alpha_1/2$ and $\tilde{\alpha}_2 = -\alpha_1/2$ and Eq.~\eqref{equ:charform} gives 
\begin{align}
    \chi_\lambda(u_c) & = \tr{e^{i\alpha_1 D^\lambda_1}}\\
    & =\det\begin{bmatrix}
         e^{i\alpha_1(\lambda_1+1)/2} & e^{i\alpha_1\lambda_2/2}\\
         e^{-i\alpha_1(\lambda_1+1)/2}& e^{-i\alpha_1\lambda_2/2}
    \end{bmatrix}
    \det\begin{bmatrix}
         e^{i\alpha_1/2} &1\\
         e^{-i\alpha_1/2} &1
    \end{bmatrix}^{-1}\nonumber\\
    & = \frac{\sin[\alpha_1(2J+1)/2]}{\sin[\alpha_1/2]},\nonumber 
\end{align}
where we have substituted $J=(\lambda_1-\lambda_2)/2$ in the second line. 
We note that this result could have been obtained directly from Eq.~\eqref{equ:charcsg} upon recalling that $D_1=S_z^J$ represents the $z$-component of a spin-$J$ system and thus has the matrix form $S^J_z= \text{diag}[J,J-1,\dots,-J]$. 
For $d>2$, however, such a direct approach is no longer feasible. 
The character formula $\eqref{equ:charform}$ then becomes a valuable tool, as we shall see in the following. 

\section{Set-up} \label{sec:setup}
Permutation-invariant systems encompass all those in which the dynamics are generated by Hamiltonians that are unchanged under exchange of any subsystems.
Here, our main objects of interest are ensembles of $n$ identical, non-interacting $d$-level systems, to which we refer as particles.
In the following, we show how the dynamics and stationary states that emerge when such an ensemble is coupled to a thermal bath can be systematically described using the tools of the previous section. 

\subsection{Multi-frequency systems} \label{sec:multi-freq}
\subsubsection{Dynamics}
We assume that the Hamiltonian of a single particle has the form 
\begin{equation}\label{equ:SPHamDef}
h=\hbar\Omega \sum_i a_i d_i,
\end{equation}
where $\Omega$ sets the overall energy scale, the $a_i\in\mathbb{R}$ are dimensionless constants and the $d_i$ are the diagonal generators of $\mathrm{SU}(d)$. 
The Hamiltonian of the ensemble is thus given by 
\begin{equation}\label{equ:Ham_setup}
H = \hbar\Omega \sum_i a_i D_i,
\end{equation}
with the $D_i$ being the diagonal generators of the tensor-product representation of $\mathrm{SU}(d)$ on the Hilbert space $\mathcal{H}=(\mathbb{C}^d)^{\otimes n}$.
Note that, from here onward, we drop the index $(n)$ that was used in Sec.~\ref{sec:SW} to denote the tensor-product representation unless it is required for clarity. 

The ensemble is coupled to a thermal bath at inverse temperature $\beta = 1/k_\text{B} T$, where $k_\text{B}$ denotes Boltzmann's constant. 
We assume that the bath cannot distinguish between the particles of the ensemble so that the system-bath coupling can be described in terms of the collective interaction Hamiltonian 
\begin{align}\label{equ:interaction_ham}
    H_{I}=\hbar \Delta \sum_\mu E_\mu \otimes B_\mu. 
\end{align}
Here, the $E_\mu$ are the non-diagonal generators of the tensor-product representation of $\mathrm{SU}(d)$, the Hermitian operators $B_\mu=B_{-\mu}$ correspond to observables of the bath and $\Delta$ sets the coupling strength. 
From the commutation relation $[D_i,E_\mu]=v^i_\mu E_\mu$, we can now determine the relevant Bohr frequencies $\omega_\mu$ of the system (i.e., the gaps in its spectrum), which are defined by the relation 
\begin{equation}
[H,E_\mu] = \hbar\Omega \sum_i a_i v_\mu^i E_\mu =\hbar\omega_\mu E_\mu. 
\end{equation}
Thus, we have $\omega_\mu= \Omega \sum_i  a_i v_\mu^i$, where $v^i_\mu$ is the element $i$ of the root vector associated with the generator $e_\mu$.

To describe the dynamics of the ensemble, we apply the standard weak-coupling, Born-Markov and secular approximations. 
These approximations require that the time scale of the system-bath interaction, which determined be the inverse coupling strength, is much larger than the relaxation time of the bath and the time scale of the bare system, which is determined by its Bohr frequencies \cite{breuer2002theory}. 
Under these conditions, one can derive the collective weak-coupling master equation
\begin{align}\label{equ:master}
\partial_t \rho_t =& -\frac{i}{\hbar}[H + H_{LS},\rho_t]\\
& + \sum_{\langle \mu,\nu \rangle} \frac{\Gamma^{\mu\nu}_{\omega}}{2}
    \Bigl\{ [E_\nu ,\rho_t E_\mu^{\dagger}] + [E_\nu \rho_t, E_\mu^{\dagger}] \Bigr\}, \nonumber
\end{align}
where $\rho_t$ denotes the state of the ensemble, the $E_\nu$ and $E_\mu^\dagger$ play the role of jump operators and the Lamb shift is given by $H_{LS}=\hbar \Delta \sum_{\langle \mu,\nu \rangle} s^{\mu \nu}_\omega E_\mu^\dagger E_\nu$, where $s^{\mu \nu}_\omega$ is the anti-Hermitian part of the bath correlation matrix~\cite{breuer2002theory}.
$H_{LS}$ commutes with the system Hamiltonian $H$ and does not enter into the steady state. 
The sum in Eq.~\eqref{equ:master} runs over all pairs of indices $\langle \mu,\nu \rangle$ for which $\omega_\mu=\omega_\nu=\omega$. 
The complex coefficients $\Gamma^{\mu\nu}_{\omega}$ are determined by the bath-correlation functions. 
They form a positive-semidefinite Hermitian matrix and obey the detailed balance condition $\Gamma^{\mu\nu}_{-\omega}=e^{\beta\hbar\omega} \Gamma^{\nu\mu}_{\omega}$ \cite{breuer2002theory}.
Here, we further assume that the bath couples to all collective modes of the system independently so that the matrix $\Gamma^{\mu\nu}_{\omega}$ has full rank for every $\omega\neq 0$. 

Schur-Weyl duality now makes it possible to simplify the dynamics of the system as follows. 
We first recall that any operator $O$ that is invariant under arbitrary particle permutations takes the block-diagonal form 
\begin{equation}\label{equ:blockobs}
O = \bigoplus_\lambda \mathbbm{1}_{\mathcal{K}^\lambda}\otimes O^\lambda
\end{equation}
in the Schur basis, where the index $\lambda$ corresponds to an ordered partition of $n$ into $d$ integers, $\mathbbm{1}_{\mathcal{K}^\lambda}$ is the identity operator on the permutation subspace $\mathcal{K}^\lambda$ and the operator $O^\lambda$ acts on the unitary subspace $\mathcal{H}^\lambda$. 
This result implies that any permutation-invariant state of the ensemble can be written as 
\begin{equation}\label{equ:statedecomp}
    \rho = \bigoplus_\lambda p^\lambda \frac{\id_{\mathcal{K}^\lambda}}{m_\lambda} \otimes \rho^\lambda . 
\end{equation}
Here, $m_\lambda = \text{dim}\;\mathcal{K}^\lambda$ is the multiplicity of the irrep $\lambda$ and we have applied the normalisation condition $\tr{\rho^\lambda_t}=1$ so that the $p^\lambda\geq 0$ add up to $1$ and can thus be regarded as the probabilities for the system to occupy the diagonal block $\lambda$. 
Since the operators $H$, $H_{LS}$ and $E_\mu$, which enter the collective master equation \eqref{equ:master}, all decompose according to Eq.~\eqref{equ:blockobs}, the block structure of the state \eqref{equ:statedecomp} is preserved at any later time. 
Thus, if the system is initially in  a permutation-invariant state, the block-occupation probabilities $p^\lambda_t = p^\lambda$ are conserved and the sub-states $\rho^{\lambda}_t$ follow the master equation \eqref{equ:master} with $H$, $H_{LS}$ and $E_\mu$ replaced by $H^{\lambda}$, $H^{\lambda}_{LS}$ and $E^{\lambda}_\mu$, respectively. 

It is now convenient to introduce the reduced state 
\begin{equation} \label{eqn:reduced_state}
	\tilde{\rho}_t = \bigoplus_\lambda p^\lambda \rho^\lambda_t, 
\end{equation}
which is obtained by tracing out the permutation subspaces $\mathcal{K}^\lambda$.
This state lives on the \emph{reduced Hilbert space} $\tilde{\mc{H}} = \bigoplus_\lambda \mc{H}^\lambda$, which contains all degrees of freedom that are available to an observer who has access only to permutation-invariant observables; 
the expectation values of such observables are given by $\tr{O\rho_t} =\sum_\lambda p^\lambda \tr{O^\lambda \rho^\lambda_t}$. 
As we will see in Sec.~\ref{sec:ThermoQuant}, the reduced state fully determines the operationally accessible thermodynamical properties of the system.
Also note that for any general state $\rho_t$, the reduced state $\tilde{\rho_t}$ must always be block-diagonal. 
That is, there is a superselection rule preventing the existence of superpositions of different $\lambda$ on this space.

\subsubsection{Steady states}
In the long-time limit $t\rightarrow\infty$, the ensemble settles to a steady state $\rho_\infty$. 
For $n>1$, this state is not unique. 
It does, however, admit a universal structure. 
In App.~\ref{app:steady_state} we prove a general theorem, which shows that, as long as the dissipative part of the master equation contains a set of simple roots, the full-rank condition on the matrix  $\Gamma^{\mu \nu}_\omega$ and the detailed-balance condition $\Gamma^{\mu\nu}_{-\omega}=e^{\beta\hbar\omega}\Gamma^{\mu\nu}_\omega$ imply that all steady states take the block diagonal form
\begin{equation}\label{equ:sscollME}
    \rho_\infty = \bigoplus_\lambda p^\lambda \sigma^\lambda \otimes \gamma^\lambda_\beta
\end{equation}
in the Schur basis.
Here, the operators $\sigma^\lambda$, which have trace $1$, and the block-occupation probabilities $p^\lambda$ depend on the initial state~\footnote{Note that, by choosing an appropriate initial state (regardless of whether pure or mixed), any set of $p^\lambda$ is realisable.} and 
\begin{equation}
    \gamma_\beta^\lambda = \frac{e^{-\beta H^\lambda}}{Z_\beta^\lambda}
\end{equation}
is a Gibbs state with respect to the irrep $H^\lambda$ of $H$, where the partial partition function $Z_\beta^\lambda$ is fixed by the condition $\tr{\gamma^\lambda_\beta}=1$. 
That is, every block thermalises independently, subject to $\lambda$ being an effective conserved quantity.
We stress that this result holds for any initial state, regardless of whether it has the block structure \eqref{equ:statedecomp}. 
Moreover, our theorem holds more generally for any permutation-invariant system Hamiltonian, which may in principle also contain interactions between particles, and even if the detailed-balance condition does not hold, in which case the Gibbs state $\gamma^\lambda_\beta$ has to be replaced with some general unique state $\rho^\lambda$ -- for details see App.~\ref{app:steady_state}. 

A natural choice for the initial state of the ensemble, on which we will focus in the following, is given by the Gibbs state 
\begin{equation}\label{equ:iniGibbs}
    \gamma_{\beta_0} = \frac{e^{-\beta_0 H}}{Z_{\beta_0}} 
    	= \bigoplus_\lambda \frac{ m_\lambda Z^\lambda_{\beta_0}}{Z_{\beta_0}} \frac{\id_{\mathcal{K}^\lambda}}{m_\lambda} \otimes \gamma^\lambda_{\beta_0}. 
\end{equation}
Here, $\beta_0 = 1/k_\text{B} T_0 $ is the inverse temperature of some environment, in which the ensemble has initially thermalised, and  $Z_{\beta_0}=z_{\beta_0}^n$, where $z_{\beta_0}=\tr{e^{-\beta_0 h}}$ is the single-particle partition function. 
The block-occupation probabilities are now given by $p^\lambda = p^\lambda_{\beta_0}= m_\lambda Z^\lambda_{\beta_0}/Z_{\beta_0}$. 
The steady state thus becomes
\begin{equation}\label{equ:SS}
	\rho_\infty = \rho_{\beta,\beta_0}
	:= \bigoplus_\lambda  p^\lambda_{\beta_0} \frac{\id_{\mathcal{K}^\lambda}}{m_\lambda} \otimes \gamma^\lambda_\beta. 
\end{equation}
Finally, tracing out the permutation subspaces in Eqs.~\eqref{equ:iniGibbs} and \eqref{equ:SS}, respectively, gives the reduced Gibbs and steady states 
\begin{equation}\label{equ:redSS} 
	\tilde{\gamma}_{\beta_0} = \bigoplus_\lambda p^\lambda_{\beta_0} \gamma^\lambda_{\beta_0},
	\quad
	\tilde{\rho}_{\beta,\beta_0} = \bigoplus_\lambda p^\lambda_{\beta_0} \gamma^\lambda_\beta.
\end{equation}
We note that, for $\beta=\beta_0$, the Gibbs state $\gamma_{\beta_0}$ is a stationary solution of the master equation \eqref{equ:master}, which implies that the ensemble remains in thermal equilibrium, i.e., $\tilde{\rho}_{\beta,\beta}= \tilde{\gamma}_{\beta}$.  
However, for $\beta\neq\beta_0$ the properties of the steady state can deviate substantially from those of a thermal state, as we will show in Sec.~\ref{sec:ThermoQuant}.  

\subsection{Spin systems}
As a reference for our results, we will consider ensembles of non-interacting spin-$s$ particles, which have been analysed in detail in Ref.~\cite{Latune2019}. 
The dynamical variables of a single spin-$s$ system, $S_{x,y,z}^s$, form an irrep of the Lie algebra $su(2)$ on the Hilbert space $\mathbb{C}^{2s+1}$. 
The corresponding tensor-product representation on the ensemble Hilbert space $\mathcal{H} = (\mathbb{C}^{2s+1})^{\otimes n}$ is given by 
\begin{equation}
    S^{s(n)}_{x,y,z} = \sum_{k=0}^{n-1} \mathbbm{1}_{2s+1}^{\otimes k}\otimes S_{x,y,z}^s\otimes\mathbbm{1}_{2s+1}^{\otimes (n-1-k)}. 
\end{equation}
Hence, with the single-particle Hamiltonian $h=\hbar\Omega S^s_z$, the ensemble Hamiltonian is $H=\hbar\Omega D_1$ with $D_1=S^{s(n)}_z$. 
Choosing the system and bath to couple instead via a collective spin,
\begin{equation} \label{eqn:interaction_spin}
    H_I = \hbar \Delta \, S^{s(n)}_x \otimes B,
\end{equation}
the Lamb shift and the jump operators in the master equation \eqref{equ:master} are then given by $H_{LS} = \hbar \Delta (s^{++}_\Omega E_- E_+ + s^{--}_{-\Omega} E_+ E_-)$ and $E_{\pm}= S^{s(n)}_x \pm i S^{s(n)}_y$.
Note that, in contrast to general $d$-level systems, spin systems are characterised by a single Bohr frequency $\omega_\pm = \pm \Omega$. 

For $s>1/2$, the irrep $S^s_{x,y,z}$ is no longer the natural representation of $su(2)$. 
Therefore, the Schur-Weyl duality does not apply in this case.  
Nevertheless, the operators $D_1$ and $E_\pm$ still take the block-diagonal form \eqref{equ:blockobs} in the collective angular momentum basis $\ket{J,m}$, where $J$ is the total angular momentum quantum number and $m$ is the magnetic quantum number. 
However, the blocks corresponding to each $J$ are subdivisions of those implied by Schur-Weyl duality.
Consequently, the results of Sec.~\ref{sec:multi-freq} equally apply to spin-$s$ particles, where the irrep index $\lambda$ has to be replaced with $J$. 
The crucial difference is that we can no longer use the formulas \eqref{equ:dimUlambda} and \eqref{equ:multformSWSpin} to calculate the dimensions and multiplicities of the individual irreps. 
Instead, the dimension of an irrep $J$ is now given by $d_J = 2J +1$ and the multiplicities have to be determined from the standard rules for the addition of angular momenta, which lead to the recursion relation 
\begin{equation}\label{equ:spinrec}
    m_J(n+1) = \sum_{\substack{J' :\; |J'-s|\leq J\leq J' +s,\\ J'+s-J \, \in\,  \mathbb{Z}}} m_{J'}(n),
\end{equation}
where $n$ is the number of particles. 
The theorem in App.~\ref{app:steady_state} also holds for spin systems, giving rise to steady states of the same form as Eq.~\eqref{equ:sscollME} but with $\lambda$ replaced by $J$.
This also settles a conjecture from Ref.~\cite{Latune2019} that the off-diagonal blocks vanish in the steady state.

\renewcommand{\ketbra}[2]{|#1\rangle\langle #2|}

\section{Thermodynamical quantities}\label{sec:ThermoQuant}
In this section, we show how the steady-state energy, reduced entropy and reduced 	non-equilibrium free energy can be  calculated for the set-up laid out in the last section.  
To illustrate our results, we compare an ensemble of spin-1 particles with an ensemble of $3$-level particles with $\mathrm{SU}(3)$ symmetry. We further derive explicit expressions for our thermodynamical quantities of interest in certain limiting cases as functions of $s$ and $d$ for spin-$s$ particles and general $d$-level particles, respectively. 

\subsection{General expressions}

For a general state $\rho$, and some fixed inverse temperature $\beta$, the steady-state energy, reduced entropy and reduced non-equilibrium free energy are defined according to the block structure~\eqref{eqn:reduced_state} as  
\begin{align}
\label{equ:EnDef}
E(\rho) & = \tr{\rho H} 
	= \sum_\lambda p^\lambda \tr{\rho^\lambda H^\lambda}
	= \sum_\lambda p^\lambda E(\rho^\lambda),\\
\label{equ:RedSDef}
\tilde{S}(\rho) & = \sum_\lambda p^\lambda S(\rho^\lambda), \\
\label{equ:RedNEFDef}
\tilde{F}(\rho) & = E(\rho) - \tilde{S}(\rho)/\beta 
	= \sum_\lambda p^\lambda \bigl(E(\rho^\lambda) - S(\rho^\lambda)/\beta\bigr),
\end{align}
where $S(\rho)= -\tr{\rho\ln[\rho]}$ denotes the von Neumann entropy.
The reduced quantities $\tilde{S}(\rho)$ and $\tilde{F}(\rho)$ have been constructed so that they do not contain entropic contributions from the degeneracy spaces $\mc{K}^\lambda$.
This approach is motivated by the assumption that all degrees of freedom that interact with the experimenter's apparatus or the environment are represented by permutation-invariant observables, which do not give access to any information stored in the subspaces $\mathcal{K}^\lambda$. 
As a result, the quantities $\tilde{S}(\rho)$ and $\tilde{F}(\rho)$ can, like $E(\rho)$, be expressed as averages over their respective counterparts on the individual irreps $\lambda$ with respect to the probability distribution $p^\lambda$. 
Note that, for steady states of the form $\bigoplus_\lambda p^\lambda \sigma^\lambda \otimes \rho^\lambda$, we can write the reduced entropy as a difference $\tilde S(\rho) = S(\rho) - S(\check{\rho})$, where $\check{\rho} = \bigoplus_\lambda p^\lambda \sigma^\lambda$ is obtained from $\rho$ by tracing out the unitary subspaces.

For the steady state \eqref{equ:SS}, the thermodynamical quantities \eqref{equ:EnDef}-\eqref{equ:RedNEFDef} become 
\begin{align}
\label{equ:SSEn}
E(\rho_{\beta,\beta_0}) & = E_{\beta,\beta_0}
	= - \sum_\lambda p^\lambda_{\beta_0}\partial_\beta \ln[Z^\lambda_\beta],\\
\label{equ:redEnt}
\tilde{S}(\rho_{\beta,\beta_0}) & = \tilde{S}_{\beta,\beta_0}
	= -\sum_\lambda p^\lambda_{\beta_0} 
\bigl(\beta\partial_\beta \ln[Z^\lambda_\beta]-\ln[Z^\lambda_\beta]\bigr),\\
\label{equ:freeEn}
\tilde{F}(\rho_{\beta,\beta_0})& = \tilde{F}_{\beta,\beta_0}
	= -\frac{1}{\beta}\sum_\lambda p^\lambda_{\beta_0} \ln[Z^\lambda_\beta]
\end{align}
with $p^\lambda_{\beta_0} = m_\lambda Z^\lambda_{\beta_0}/Z_{\beta_0}$. 
To evaluate these expressions, we have to calculate the partial partition functions $Z^\lambda_{\beta}$, the total partition function $Z_\beta$ and the multiplicities of the individual irreps $m_\lambda$. 

To this end, we first observe that $Z^\lambda_{\beta}$ can be written as 
\begin{equation}\label{equ:PartZ}
Z^\lambda_{\beta}	= \tr{e^{-\beta\sum_i a_i D^\lambda_i}}
					= \tr{e^{i\sum_i \alpha_i D^\lambda_i}},
\end{equation}
where we have replaced $a_i$ by $-i\alpha_i/\beta_0$ in the second expression. 
Since the $D^\lambda_i$ are diagonal generators of an irrep of $\mathrm{SU}(d)$, this quantity can be calculated by means of analytic continuation of Weyl's character formula, which we have discussed in Sec.~\ref{sec:characters}. 
For $d=3$, we find~\cite{greiner2012quantum}
\begin{align}\label{equ:partZSU3}  
Z^\lambda_{\beta}  & =  e^{\beta a_2(2x_1 + x_2)/3}\\
	& \times \sum_{k=0}^{x_1} \sum_{l=0}^{x_2}
	e^{-\beta a_2 (k+l)}
	\frac{\sinh[\beta a_1 (k-l+x_2 +1)/2]}{\sinh[\beta a_1/2]},\nonumber
\end{align}
where $x_j= \lambda_j - \lambda_{j+1}$ and $\lambda$ denotes the ordered partition $(\lambda_1,\lambda_2,\lambda_3)$. 
Analogous formulas can in principle be derived for any larger $d$. 
The single-particle partition function $z_{\beta}$ is obtained as a special case for $\lambda=(1,0,\dots,0)$. 
The total partition function is then given by $Z_{\beta} = (z_{\beta})^n$. 
For an ensembles of spin-$s$ systems the partial partition function over the irrep $J$ is given by 
\begin{equation}\label{equ:partZSU2}
Z^J_\beta = \frac{\sinh[\beta(2J+1)/2]}{\sinh[\beta/2]}.
\end{equation}
as can be easily verified by direct computation. 

If the individual particles are described in terms of the natural representation of $\mathrm{SU}(d)$, the multiplicities $m_\lambda$ can be obtained from the combinatorial formula \eqref{equ:multformSW}, which yields  
\begin{equation}
m_\lambda = \frac{n!(\tilde{\lambda}_1-\tilde{\lambda}_2)
		(\tilde{\lambda}_1-\tilde{\lambda}_3)
		(\tilde{\lambda}_2-\tilde{\lambda}_3)}{
	\tilde{\lambda}_1!\tilde{\lambda}_2!\tilde{\lambda}_3!}
\end{equation}
for $d=3$ with $\tilde{\lambda}_j$ being defined in Eq.~\eqref{equ:dimUlambda}. 
For spin systems with $s>1/2$, the multiplicities can be obtained from the recursion relation \eqref{equ:spinrec}. 
With these prerequisites, the quantities \eqref{equ:SSEn}-\eqref{equ:freeEn} are computationally accessible. 

\subsection{$\mathrm{SU}(2)$ vs $\mathrm{SU}(3)$}

\begin{figure}[h!]
    \centering
    \includegraphics[width=.45\textwidth]{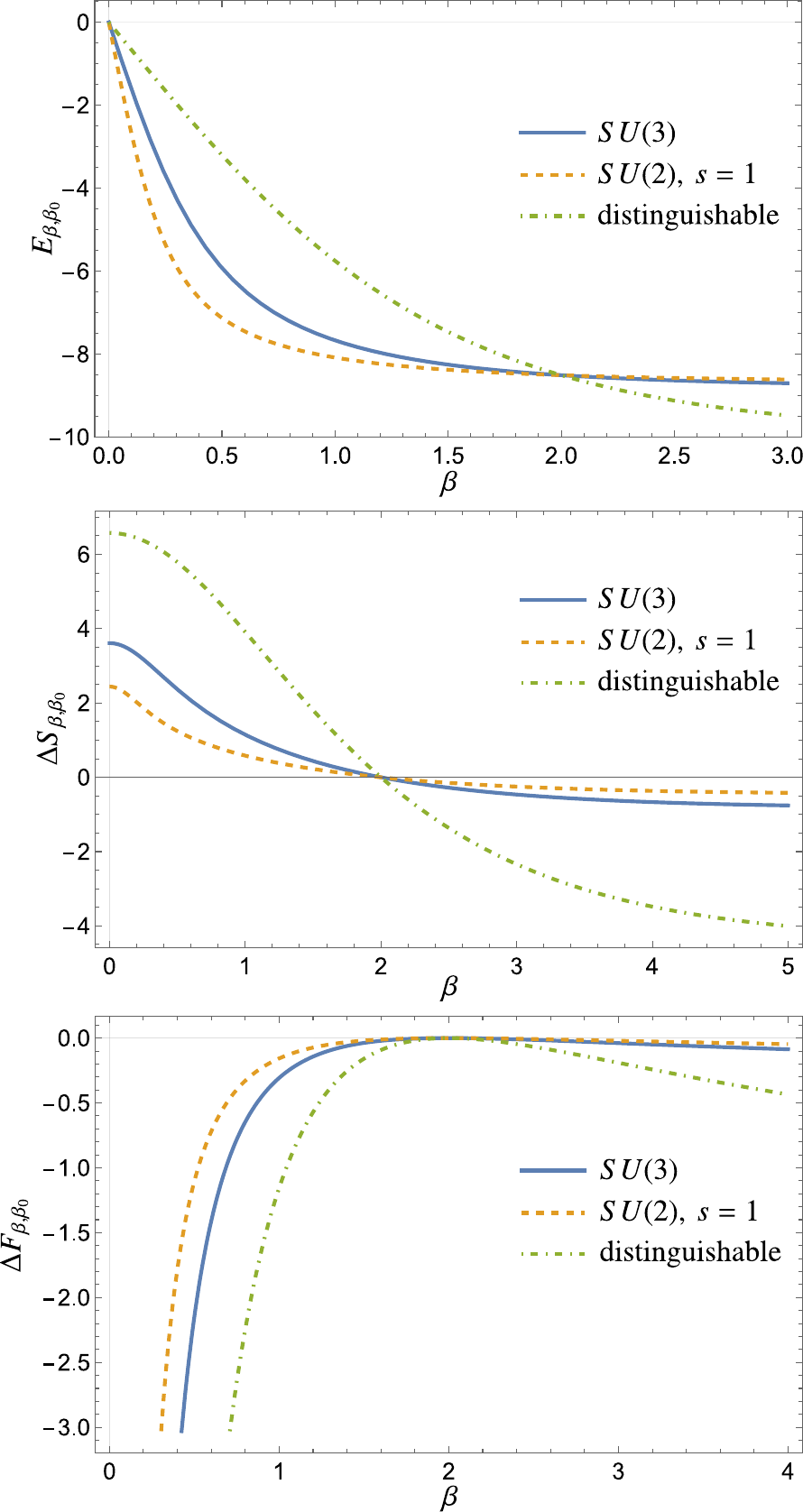}
    \caption{
    Plots of the steady-state energy \eqref{equ:SSEn}, the reduced entropy change \eqref{equ:redEnt} and the change in the reduced free energy \eqref{equ:FreeEnergyChange} (top to bottom) as a functions of the inverse bath temperature $\beta$ for ensembles of $3$-level systems with $\mathrm{SU}(3)$- and $\mathrm{SU}(2)$-symmetry, starting from the thermal state \eqref{equ:iniGibbs} at $\beta_0$.  
    For comparison, the dot-dashed lines labelled ``distinguishable" show the fully thermal state energy $E_{\beta,\beta}$ (top), and the changes in entropy (middle) and free energy (bottom) during a full equilibration process of distinguishable particles from $\beta_0$ to $\beta$. 
    For all plots, we have chosen the single-particle Hamiltonian \eqref{equ:SPHam} and set $n=10$ and $\beta_0=2$.
    Note that, for $\beta=\beta_0$, the initial Gibbs state $\gamma_{\beta_0}$ is a stationary state of the master equation \eqref{equ:master}. 
    Therefore, all curves intersect at $\beta=2$, and all entropy and free energy differences are 0 at $\beta=2$ in the lower two plots.}
    \label{fig:Energy}
\end{figure}

As a first application of our theory, we investigate how the transition from $\mathrm{SU}(2)$ to $\mathrm{SU}(3)$ changes the thermodynamical properties of the steady state. 
For convenience, we now set Boltzmann's constant to $1$ and rescale all energies and temperatures with $\hbar\Omega$ so that $\beta$, $\beta_0$ and $h,H, H^\lambda$ become dimensionless from here onward.

As the arguably simplest setting in which $\mathrm{SU}(2)$ and $\mathrm{SU}(3)$ can be compared, we consider ensembles of 3-level systems with excited, intermediate and ground states $\ket{+}, \ket{0}$ and $\ket{-}$. 
To isolate the effect of the symmetry group, we pick the single-particle Hamiltonian 
\begin{equation}\label{equ:SPHam}
	h= \dyad{+} - \dyad{-}. 
\end{equation}
In an $\mathrm{SU}(2)$ description, this Hamiltonian corresponds to a spin-1 system with $h= d_1 = S^1_z$, where $S^1_z$ was defined in Eq.~\eqref{equ:s1Pauli}. 
In terms of $\mathrm{SU}(3)$, the Hamiltonian \eqref{equ:SPHam} describes a ladder system with $h= (d_1 + \sqrt{3} d_2)/2 = (\Lambda_3 + \sqrt{3}\Lambda_8)/2$, where we have identified the diagonal generators $d_1$, $d_2$ with the Gell-Mann matrices from Eq.~\eqref{equ:GellMann}.

Although the single-particle Hamiltonian is identical in both cases, the steady state \eqref{equ:redSS} is not the same for $\mathrm{SU}(2)$ and $\mathrm{SU}(3)$, since different system-bath couplings (Eqs.~(\ref{eqn:interaction_spin}),(\ref{equ:interaction_ham})) lead to different dissipation mechanisms. 
For $\mathrm{SU}(2)$, the ensemble relaxes via a single dissipation channel described by the Lindblad operators $E_\pm$, which represent the non-diagonal generators 
\begin{equation}
e_+ = e_-^\dagger = \frac{1}{\sqrt{2}} \bigl( \ketbra{+}{0} + \ketbra{0}{-}\bigr).
\end{equation}
By contrast, for $\mathrm{SU}(3)$, relaxation to the steady state occurs via $3$ dissipation channels, whose Lindblad operators $E_{1\pm}, E_{2\pm}$ and $E_{3\pm}$ represent the generators 
\begin{align}
& e_{1+}=e_{1-}^\dagger = \ketbra{+}{0}, \quad e_{2+}=e_{2-}^\dagger = \ketbra{+}{-},\\
& e_{3+}=e_{3-}^\dagger = \ketbra{0}{-}. \nonumber
\end{align}
Hence, in the $\mathrm{SU}(2)$-case the Lindblad operators induce coherent superpositions of jumps between intermediate and excited and ground and intermediate states, while for $\mathrm{SU}(3)$ all possible transitions in the single-particle system are addressed separately by the bath. 

This difference changes the structure of the steady state and thus its thermodynamical properties. 
In Fig.~\ref{fig:Energy}, we plot the internal energy $E_{\beta,\beta_0}$, and changes in reduced entropy and reduced free energy during the relaxation process, which are given by 
\begin{align}\label{equ:FreeEnergyChange}
    \Delta \tilde{S}_{\beta,\beta_0} & = \tilde{S}_{\beta,\beta_0} - \tilde{S}_{\beta_0,\beta_0}, \nonumber \\
    \Delta  \tilde{F}_{\beta,\beta_0} & = \tilde{F}_{\beta,\beta_0} - \tilde{F}_{\beta_0,\beta_0}.
\end{align}
From these plots, we immediately note two features. 
First, as expected, all three quantities deviate from their thermal equilibrium references in the steady state \eqref{equ:SS}. 
Second, although we use the same single-particle Hamiltonian for both symmetry groups, these deviations are typically stronger for $\mathrm{SU}(2)$ than for $\mathrm{SU}(3)$. 
This result suggest that the 3-channel dissipation mechanism, which applies to systems with $\mathrm{SU}(3)$-symmetry, allows the system to come closer to thermal equilibrium than the single-channel mechanism applying to spin systems.

\subsection{Limiting cases}
We now return to the general case of ensembles with $\mathrm{SU}(d)$-symmetry. 
In the following, we derive explicit expressions for the thermodynamical quantities \eqref{equ:SSEn}-\eqref{equ:freeEn} as functions of $d$ in special limits to show how our theory can be applied in practice. For comparison, we also analyse spin-$s$ ensembles. 

\subsubsection{Low initial and high bath temperature}
In the limit $\beta_0 \gg 1$, the ensemble will almost exclusively occupy the trivial symmetric subspace, which is spanned by permutation-invariant many-particle states. 
To understand this effect, we consider the probability $p^g_{\beta_0} = \bra{g^n}\gamma_{\beta_0}\ket{g^n}$ of finding the ensemble in its non-degenerate ground state $\ket{g^n}=\otimes_{k=1}^n \ket{g}^{(k)}$, where $\ket{g}^{(k)}$ is the ground state of the particle $k$. 
Since the state $\ket{g^n}$ belongs to the symmetric subspace, the probability of finding the system in this subspace is subject to the bounds $1\geq p^\text{sym}_{\beta_0}\geq p^g_{\beta_0}$. 
Taking the low-temperature limit $\beta_0 \rightarrow\infty $ gives $p^g_{\beta_0}=1$ and thus $p^\text{sym}_{\beta_0}=1$. 
Hence, by continuity, we have $p^\text{sym}_{\beta_0}\simeq 1$ for $\beta_0\gg 1$. 
Since the block-structure \eqref{equ:iniGibbs} of the state $\gamma_{\beta_0}$ is conserved by the master equation \eqref{equ:master}, so is the probability $p^\text{sym}_{\beta_0}$. 
Consequently, the steady state of the ensemble is also nearly confined to the symmetric subspace and the thermodynamical quantities \eqref{equ:SSEn}-\eqref{equ:freeEn} are dominated by contributions from the symmetric irrep $\lambda_\text{sym}=(n,0,\dots,0)$. 
This observation makes it possible to derive asymptotically exact expressions for these quantities in the limit $\beta \ll 1$. 
That is, we consider a situation where the system is initially thermalised in a low-temperature environment and then brought into contact with a high-temperature bath. 

For an $\mathrm{SU}(d)$-ensemble with single-particle Hamiltonian $h=\text{diag}[\varepsilon_1,\dots,\varepsilon_d]$, the partial partition function $Z^\text{sym}_\beta$ of the symmetric subspaces is given by 
\begin{equation}\label{equ:PPFd}
Z^\text{sym}_\beta = d_\text{sym} \left(1 +\beta^2 \frac{n(n+d)}{2(d+1)}\llangle \varepsilon^2 \rrangle\right)
+ \mathcal{O}(\beta^3)
\end{equation}
as we prove in App.~\ref{app:steady_state_sym_E}. 
Here,  
\begin{equation}
d_\text{sym} = \binom{n+d-1}{n}
\end{equation}
is the dimension of the symmetric subspace, which can be obtained from Eq.~\eqref{equ:dimUlambda}, and 
\begin{equation}
\llangle \varepsilon^2\rrangle = \frac{1}{d}\sum_{i=1}^d \varepsilon_i^2
\end{equation}
is the variance of the single-particle energy at infinite temperature.
Note that we have $\langle\varepsilon\rangle = \frac{1}{d}\sum_{i=1}^d \varepsilon_i = 0$, since the single-particle Hamiltonian is traceless by assumption. 
Using the expansions \eqref{equ:PPFd} and \eqref{PPFs}, we find that the quantities \eqref{equ:SSEn}-\eqref{equ:freeEn} become 
\begin{align}
\label{equ:CorrE}
E_{\beta,\infty} &=  -2\beta q^d_n +\mathcal{O}(\beta^2),\\
\tilde{S}_{\beta,\infty} &= \ln[d_\text{sym}]-\beta^2 q^d_n + \mathcal{O}(\beta^3),\\
\label{equ:CorrF}
\tilde{F}_{\beta,\infty} &= -\ln[d_\text{sym}]/\beta - \beta q^d_n  +\mathcal{O}(\beta^2)
\end{align}
in the limit $\beta_0\rightarrow\infty$ with
\begin{equation}\label{equ:qsym}
q^d_n = \frac{n(n+d)\llangle \varepsilon^2 \rrangle }{2(d+1)}. 
\end{equation}

We now consider an ensemble of spin-$s$ particles.
By a similar argument to above, since the ground state belongs to the subspace of maximal angular momentum $J = ns$, the thermal state of such an ensemble will almost exclusively occupy this subspace in the limit $\beta_0 \gg 1$.
The dimension of this subspace is given by $d_\text{max}= 2ns +1$, which is strictly smaller than the dimension of the symmetric subspace for $s>1/2$ and $n>1$. 
With the single-particle Hamiltonian $h= S^s_z$, the corresponding partial partition function becomes~\cite{Latune2019}
\begin{equation}\label{PPFs}
Z^{ns}_\beta = (1+ 2ns)\left(1 +\beta^2 \frac{ns (1+ns)}{6}\right) + \mathcal{O}(\beta^3)
\end{equation}
as can be seen by expanding Eq.~\eqref{equ:partZSU2}. 
The asymptotic expressions for the internal energy, reduced entropy and reduced free energy can now be obtained by replacing $d_\text{sym}$ with $d_\text{max}$ and $q^d_n$ with 
\begin{equation}\label{equ:qspn}
Q^s_n = \frac{ns (1+ns)}{6} = \frac{n(d-1)(2+n(d-1))}{24}
\end{equation}
in Eqs.~\eqref{equ:CorrE}-\eqref{equ:CorrF}. 
Here, we have expressed $s=(d-1)/2$ in terms of the dimension $d$ of the single-particle Hilbert space to facilitate the comparison between $\mathrm{SU}(d)$- and spin-$s$ ensembles.

The expressions \eqref{equ:qsym} and \eqref{equ:qspn} show that, for the $\mathrm{SU}(d)$ ensemble as well as for the spin-$s$ ensemble, the leading-order corrections in $\beta$ to the internal energy, reduced entropy and reduced free energy of the steady state all feature a quadratic dependence on $n$ for all $d$. 
For a quantitative comparison, we choose a ladder Hamiltonian with eigenvalues $\varepsilon_i = -(d+1)/2 +i $ for the $\mathrm{SU}(d)$-ensemble, which yields $\llangle \varepsilon^2 \rrangle = (d-1)(d+1)/12$. 
We then have
\begin{equation}
q^d_n \sim n^2 (d-1)/24 \quad\text{and}\quad Q^s_n \sim n^2(d-1)^2/24
\end{equation}
for $n\gg 1$. 
Hence, in the limit of many particles, the magnitudes of the first order corrections in Eqs.~\eqref{equ:CorrE}-\eqref{equ:CorrF} are suppressed by a factor $(d-1)$ in $\mathrm{SU}(d)$-ensemble compared to the spin-$s$-ensemble. 
We stress that this effect arises solely from the different dimensions of the primarily occupied subspaces, which have multiplicity $1$ in both cases. 

\subsubsection{High initial temperature and many particles}
A second interesting limit is realised when the ensemble is initially prepared in a high-temperature state, that is, $\beta_0 \to 0$.
In the following, we calculate the steady-state energy and reduced entropy in this limit, respectively for thermalisation with a low and a high-temperature bath. 
To this end, we first observe that the probability distribution $p^\lambda_{\beta_0}$ tends to
\begin{align}
    p^\lambda_{\beta_0} \rightarrow \frac{m_\lambda Z^\lambda_{\beta_0}}{Z_{\beta_0}} = \frac{m_\lambda d_\lambda}{d^n}
\end{align}
for $\beta_0\rightarrow 0$. 
This quantity is known as a \emph{Plancherel-type measure} \cite{borodin2007asymptotics}. 
In the limit of many particles, this measure can be determined explicitly through methods of asymptotic representation theory. 
Specifically, upon changing variables from $\lambda_i$ to $\zeta_i =(\lambda_i - n/d)/\sqrt{n}$, the limiting measure for $n\rightarrow\infty$ becomes the function \cite{kerov1986asymptotic,postnova2020multiplicities}
\begin{align}\label{equ:kerov_limits}
    \phi_d(\zeta) =\frac{d^{\frac{d(d-1)+1}{2}}\left(\frac{d}{2\pi}\right)^{\frac{d-1}{2}}}{1!2!...(d-1)!}\prod_{i< j}(\zeta_i-\zeta_j)^2 e^{-\frac{d}{2}\sum_k \zeta_k^2}.
\end{align}

One interesting question is how far the mean energy of an initially hot ensemble can be reduced by thermalising with a cold bath.
When the bath temperature is low, i.e., if $\beta \to \infty$, the partial thermal state $\gamma^\lambda_\beta$ becomes a projector on the ground state within the unitary subspace $\mathcal{H}^\lambda$. 
As we show in App.~\ref{app:ground_state}, each subspace has a unique ground state with energy $\sum_{i=1}^d \lambda_i \varepsilon_i$, where $\varepsilon_i$ are the single-particle energies ordered increasingly. 
Here, we assume that $\varepsilon_1$ is non-degenerate. 
Putting this observation together with the limiting distribution \eqref{equ:kerov_limits}, we obtain the mean energy
\begin{equation} \label{eqn:many_particle_energy}
    E_{\infty,0} \rightarrow \int \dd \zeta\; \phi_d(\zeta) \, \left[\sum_{i=1}^d \lambda(\zeta)_i \varepsilon_i \right]
\end{equation}
for $n\rightarrow\infty$. 
The integral in this expression runs over all $\zeta_1,\dots,\zeta_d\in\mathbb{R}$ subject to the constraints $\zeta_1 \geq \zeta_2 \geq \cdots \geq \zeta_d$ and $\sum_{i=1}^d \zeta_i = 0$.

If we choose a ladder Hamiltonian with energy levels $\varepsilon_i = -(d+1)/2+i$ for the single-particle system, this mean energy takes the form 
\begin{equation}\label{equ:mp_ladder_energy}
E_{\infty,0} \rightarrow - \mc{E}_d \sqrt{n}
\end{equation}
for $n\rightarrow\infty$. 
Notably, this result shows that the steady-state energy of the ensemble is sub-extensive in the particle number $n$. 
Hence, thermalisation is strongly inhibited by the symmetry constraints of the system;
if the system were to instead fully thermalise to its ground state, its steady-state mean energy would be trivially given by $E_{\infty,\infty}= n \varepsilon_1 = -n(d-1)/2$. 

\begin{figure}[h!]
    \centering
    \includegraphics[width=.45\textwidth]{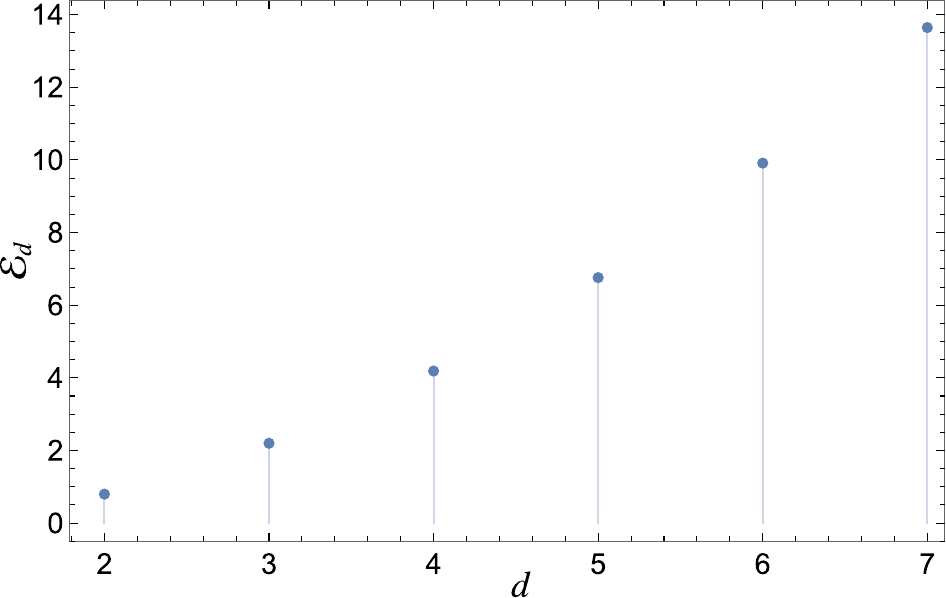}
    \caption{Plot of the limiting energy coefficient $\mc{E}_d$ as defined in Eq.~\eqref{equ:mp_ladder_energy} as a function of $d$.}
    \label{fig:Energy_asym}
\end{figure}

For $d=2$ and $d=3$, the constant $\mathcal{E}_d$ in Eq.~\eqref{equ:mp_ladder_energy} takes the  values $\mc{E}_2 = \sqrt{2/\pi}$ and $\mc{E}_3 = 9\sqrt{3/16\pi}$, see App.~\ref{app:asymptotics} for details. 
For larger dimensions, $\mathcal{E}_d$ must be determined numerically by solving the integral in Eq.~\eqref{eqn:many_particle_energy}. 
These results are plotted in Fig.~\ref{fig:Energy_asym}, which shows a non-linear increase of $\mathcal{E}_d$ with $d$ for $d\leq 7$. 
For completeness, we may again compare the $\mathrm{SU}(3)$-ensemble with a spin-1 ensemble, for which we find $E_{\infty,0} \rightarrow - \sqrt{16/3\pi}\sqrt{n}$, see App.~\ref{app:spin_asym}.
Thus, low-temperature thermalisation through the $3$-channel mechanism of the $\mathrm{SU}(3)$-case leads to a lower steady-state energy than single-channel mechanism of the spin-1 case by a factor of $27/16\approx 1.69$.
In line with our earlier results, this observation quite naturally suggests that a larger number of dissipation channels allows the system to relax closer to a thermal state. 
Note however, that the scaling of the steady-state energy with $n$ is insensitive to the dissipation mechanism. 

We now assume that the bath temperature is also high, i.e., $\beta \to 0$. 
The ensemble then remains in a high-temperature equilibrium state, whose reduced entropy is given by  
\begin{equation}
    \tilde{S}_{0,0} = \sum_\lambda p_0^\lambda \ln [d_\lambda].
\end{equation}
By using the distribution \eqref{equ:kerov_limits} for the many-particle limit, we prove in App.~\ref{app:asymptotics} that
\begin{align} \label{eqn:entropy_asym}
    \tilde{S}_{0,0} \rightarrow \int \dd \zeta\; \phi_d(\zeta) \, \ln [d_{\lambda(\zeta)}] \simeq \frac{d(d-1)}{4} \ln [n]
\end{align}
to leading order in $n$ for $n\rightarrow\infty$. 
Thus, for $d=3$, we find $\tilde{S}_{0,0} \rightarrow (3/2)\ln [n]$. 
In comparison, for a spin-1 ensemble, the reduced entropy is suppressed by a factor of $3$, that is  we have $\tilde{S}_{0,0}\rightarrow (1/2)\ln [n]$ to leading order in $n$. 
More generally, for a spin-$s$ ensemble, it is clear that $\tilde{S}_{0,0}$ is upper-bounded by the contribution $S(\gamma_{\beta}^{ns})=\ln[2ns+1] \simeq \ln [n]$ from the maximal angular momentum subspace corresponding to $J=ns$, cf. Eq.~\eqref{equ:redEnt}. 
It follows that, compared to spin ensembles, the reduced entropy $\tilde{S}_{0,0}$ of $\mathrm{SU}(d)$-ensembles is asymptotically enhanced by a factor that is at least quadratic in the dimension of the single-particle Hilbert space.

\newcommand{\bc}{\beta_\text{c}}
\newcommand{\bh}{\beta_\text{h}}

\section{Engine Cycles} \label{sec:engine}

So far we have analysed how steady-state thermodynamical quantities are affected by the symmetries of permutation-invariant ensembles of $n$ indistinguishable and non-interacting particles. 
As a next step, we explore in this section what role these symmetries play in thermodynamical processes. 
To this end, we consider a thermodynamical engine cycle that uses a collective working medium and investigate how the properties of this medium affect its performance. 

\subsection{Protocol and output}

For simplicity, we focus on the standard quantum Otto cycle \cite{Quan2007,Quan2009}.
We begin by stating the 0+4 strokes that the system undergoes during this cycle. 
In the zeroth stroke, the ensemble is initialised in the Gibbs state at the inverse temperature $\beta_0$, which determines the occupation probabilities $p^\lambda_{\beta_0}$ of the individual irreps, cf. Eq.~\eqref{equ:iniGibbs}. 
The ensemble then cyclically undergoes the four strokes of the Otto cycle between a hot and cold bath at inverse temperatures $\bh$ and $\bc$: 
\begin{itemize}
\item[(1)] Equilibration with the hot bath, 
\item[(2)] Instantaneous change of the Hamiltonian $H\rightarrow H'$,
\item[(3)] Equilibration with the cold bath, 
\item[(4)] Instantaneous change of the Hamiltonian $H'\rightarrow H.$
\end{itemize}

The net extracted work per cycle $W$, which is our main quantity of interest, is given by the sum of the work contributions from the two instantaneous strokes, 
 \begin{align}
    W=-\Tr[\rho_{\bh,\beta_0}(H'-H)]-\Tr[\rho'_{\bc,\beta_0}\left(H-H'\right)].
\end{align}
Here, $\rho_{\bh,\beta_0}$ and $\rho'_{\bc,\beta_0}$ are the steady states of the system with Hamiltonian $H$ and $H'$, respectively, see Eq.~\eqref{equ:SS}.
We now assume that the instantaneous strokes are simple compression, that is $H'=\kappa H$, where the compression factor $\kappa$ must obey the inequalities $\bh/\bc\leq\kappa\leq 1$ to ensure that a positive amount of work is generated.  
Under these conditions, the net work extraction with a collective medium becomes 
\begin{align}\label{equ:work_col}
    W^{\text{col}} & = (1-\kappa) \tr{H(\rho_{\bh,\beta_0}- \rho_{\kappa\bc,\beta_0})}\\ 
    &= (1-\kappa)(E_{\bh,\beta_0}-E_{\kappa\bc,\beta_0}) \nonumber
\end{align}
with the internal energies given by Eq.~\eqref{equ:SSEn}. 
To uncover the role of collective effects, we compare this quantity with the work generated in the same cycle with an ensemble of distinguishable particles, which fully thermalises during the isochoric strokes,
\begin{align}\label{equ:work_dis}
    W^{\text{dis}} &= (1-\kappa)\tr{H(\gamma_{\bh}-\gamma_{\kappa\beta_c})}\\ 
    & = (1-\kappa)(E_{\bh,\bh} - E_{\kappa\bc,\kappa\bc}). \nonumber 
\end{align}

\subsection{Collective work enhancement}

\begin{figure}[h!]
    \includegraphics[width=.45\textwidth]{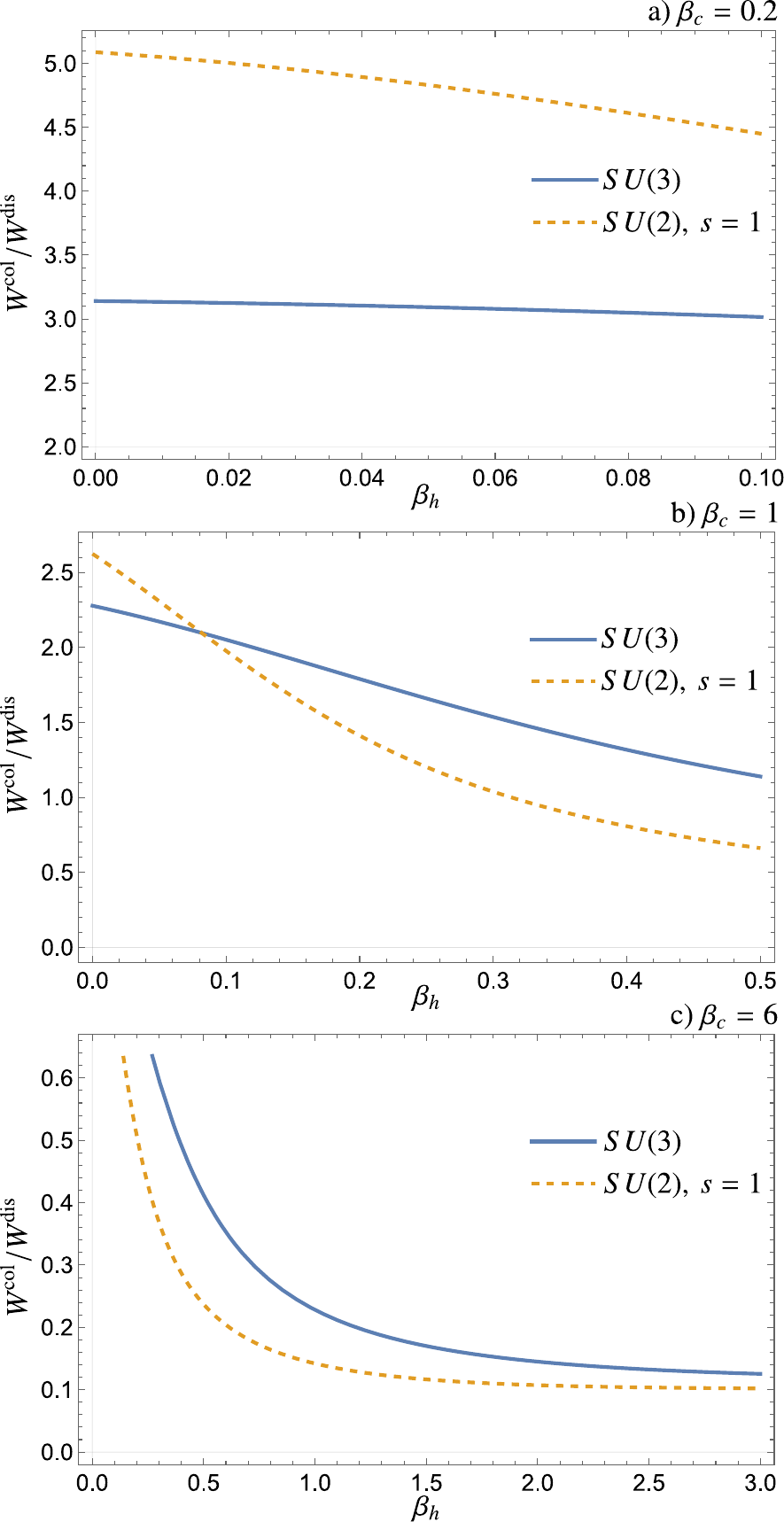}
    \caption{Plots of the work advantage $W^\text{col}/W^\text{dis}$ as a function of $\bh$ for different values of $\bc$. 
    Dashed lines correspond to spin-$1$ ensembles and solid lines to $\mathrm{SU}(3)$-ensembles. 
    For all plots, we have set $n=10$. Bath temperatures adhere to the range $\bh/\bc\leq\kappa\leq 1$, where the compression factor is $\kappa=1/2$ and the initial inverse temperature of the ensembles is $\beta_0=5$.}\label{fig:work_otto}
\end{figure}
 
Two natural questions arise at this point.
First, whether it is possible to extract more work from the collective medium than from the distinguishable one, that is, can the ratio $W^\text{col}/W^\text{dis}$ become larger than 1?
Second, whether there is an advantage in moving from spin-ensembles to ensembles with higher-order symmetry. 

Both of these questions can be answered with the help of Fig.~\ref{fig:work_otto}, where we plot the ratio $W^\text{col}/W^\text{dis}$ against $\bh$ for $\mathrm{SU}(2)$- and $\mathrm{SU}(3)$-symmetric ensembles. 
To enable a quantitative comparison, we have chosen the same single-particle Hamiltonian $h=\text{diag}[1,0,-1]$ for both cases. 
We find that, first, there is indeed a temperature range for which $W^\text{col}/W^\text{dis}>1$, meaning that more work can be extracted with the collective medium than with the distinguishable one; a similar result was found in Ref.~\cite{Latune2019} for spin-ensembles.
Second, Fig.~\ref{fig:work_otto}(b) shows a regime where $W^\text{col}/W^\text{dis}>1$ while the extracted work is larger for the $\mathrm{SU}(3)$- than for the spin-ensemble. 
An advantage of $\mathrm{SU}(3)$ over $\mathrm{SU}(2)$ is also seen in Fig.~\ref{fig:work_otto}(c), where $\kappa \bc \simeq \beta_0$. 
Here, however, $W^\text{col}/W^\text{dis}<1$, so the higher-order symmetry group rather mitigates the disadvantageous collective effects. 

The first observation can be understood in the limiting regime where the initial temperature is low and both baths temperatures are high, that is, $\beta_0 \gg 1$ and $\bh, \kappa \bc \ll 1$.
Under these conditions, we can use the asymptotic result \eqref{equ:CorrE} for the steady-state internal energy of an $\mathrm{SU}(d)$-ensemble to evaluate the work extraction from the collective medium. 
Using the formulas \eqref{equ:work_col} and \eqref{equ:work_dis}, we find
\begin{align} \label{eqn:work_ratio}
    \frac{W^\text{col}}{W^\text{dis}} \simeq \frac{n+d}{d+1}. 
\end{align}
This result shows that the work advantage grows monotonically in the number of particles $n$, but decreases monotonically with the order of the symmetry group $d$ for $n>1$.
For the special case $n=10$ and $d=3$, we recover the limiting value $W^\text{col}/W^\text{dis}\simeq 3.25$ seen in Fig~\ref{fig:work_otto}(a). 
For comparison, the analogous result for spin-$s$-ensembles derived in Ref.~\cite{Latune2019} is
\begin{align}
    \frac{W^\text{col}}{W^\text{dis}} \simeq \frac{ns+1}{s+1} = \frac{n(d-1)+2}{d+1}.
\end{align}

\begin{figure}
    \centering
    \includegraphics[width=.45\textwidth]{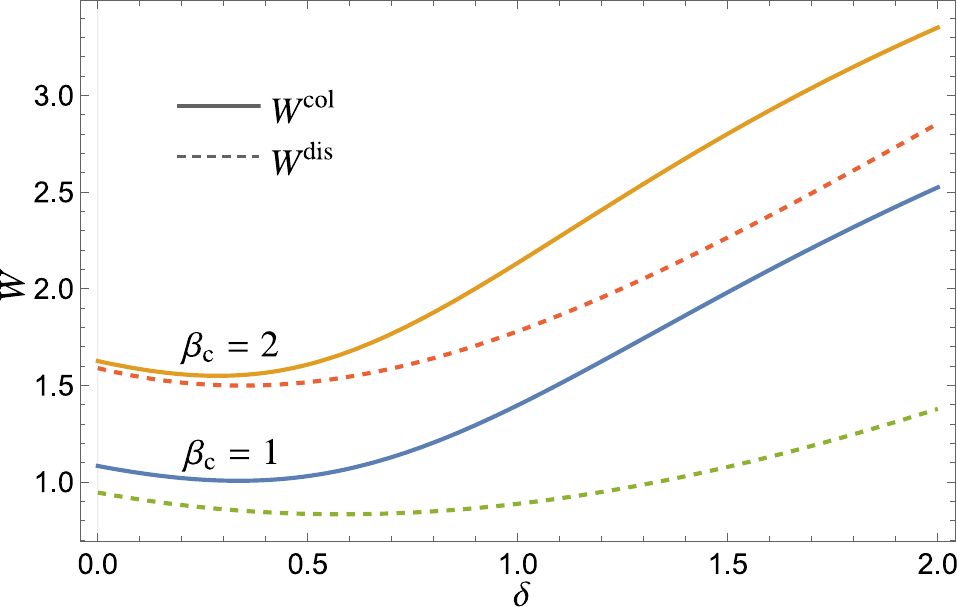}
    \caption{Plot of the work output of a quantum Otto cycle using a working medium with single-particle Hamiltonian \eqref{eqn:h_parameterised} as a function of the control parameter $\delta$. 
    Solid lines show the $\mathrm{SU}(3)$ collective ensemble and dashed lines correspond to the distinguishable ensemble, which fully thermalises in the equilibration strokes. 
    For this plot, we have set $n=7$, $\beta_0=3$, $\bh=0.1$, $\kappa=1/2$ and two different values $\bc=1,\,2$.}
    \label{fig:WorkVsA2}
\end{figure}

\subsection{Optimisation}

Higher-dimensional symmetry groups offer larger freedom in modeling the single-particle Hamiltonian as additional diagonal generators become available in the corresponding Lie algebra, cf. Eq.~\eqref{equ:SPHamDef}. 
We will now explore how this freedom can be exploited to optimise the work output of a collective quantum Otto cycle. 
To this end, we let the single-particle energies $\varepsilon_i$ vary within a bounded range, which we choose as $\varepsilon_\text{max}-\varepsilon_\text{min}=d-1$ to match a ladder Hamiltonian with unit spacing.

By using the expansion \eqref{equ:CorrE}, we can find the optimal spectrum in the limit of low initial temperatures, $\beta_0 \to \infty$, and high bath temperatures, $\bh,\kappa\bc\ll 1$. 
As shown in App.~\ref{app:steady_state_sym_E}, for even $d$, the optimal spectrum has two levels, which are $d/2$-fold degenerate with the maximal gap of $d-1$; 
that is, we have $\varepsilon_1,\dots\varepsilon_{d/2}=-(d-1)/2$ and $\varepsilon_{d/2+1},\dots,\varepsilon_d = (d-1)/2$.  For this spectrum, the optimal work output of a collective cycle and the work output generated by an ensemble of distinguishable particles are  
\begin{align}
    W^{\mathrm{col}\ast} & \simeq (1-\kappa)(\kappa \bc - \bh) \frac{n(n+d)(d-1)^2}{4(d+1)},\\
    W^{\text{dis}\ast}   & \simeq \frac{d+1}{n+d} W^{\text{col}\ast}. 
\end{align}
Hence, the work-advantage ratio is still given by Eq.~\eqref{eqn:work_ratio}. 
For odd $d$, there is a small correction, since the remaining energy level can be either at $0$ or $d-1$.

For arbitrary temperatures, we approach the optimisation problem numerically for $\mathrm{SU}(3)$-ensembles.
Normalising the difference between ground and excited single-particle energies to $2$, we can parameterise the single-particle Hamiltonian by the gap $\delta$ between the ground and first excited levels.
That is, we make the ansatz
\begin{align} 
\label{eqn:h_parameterised}
    h_\delta & = \frac{2}{3} \mathrm{diag} \left[ 2 - \frac{\delta}{2}, \delta - 1, -1 - \frac{\delta}{2} \right]\\
    &= 2(2-\delta) d_1 + \frac{2 +\delta}{2\sqrt{3}} d_2, \qquad \delta \in [0,2]. \nonumber
\end{align}
Here, we have identified the diagonal generators with the Gell-Mann matrices \eqref{equ:GellMann}, i.e., $d_1 = \Lambda_3$ and $d_2= \Lambda_8$. 
As shown in Fig.~\ref{fig:WorkVsA2}, the maximum work output is attained with a doubly degenerate excited level, corresponding to $\delta = 2$. 
Notably, the collective medium outperforms the distinguishable one for any value of $\delta$.

Our results show that the larger freedom in the choice of the single-particle Hamiltonian, which comes with higher-order symmetry groups, can indeed be used to optimise the performance of many-body Otto cycles, in addition to the boost coming from a collective dissipation mechanism.
This can be done, at least in principle, even outside the regime of extreme initial and bath temperatures. 
Furthermore, our analysis seems to suggest that the optimal single-particle Hamiltonian for collective quantum Otto cycles should feature two levels with approximately balanced degeneracy. 
Corroborating this presumption with a more systematic investigation is beyond the scope of our present work, but provides an attractive problem for future research.

\section{Higher-order symmetries: A closer look} \label{sec:higher-order}

 \begin{figure}
    \centering
    \includegraphics[width=.45\textwidth]{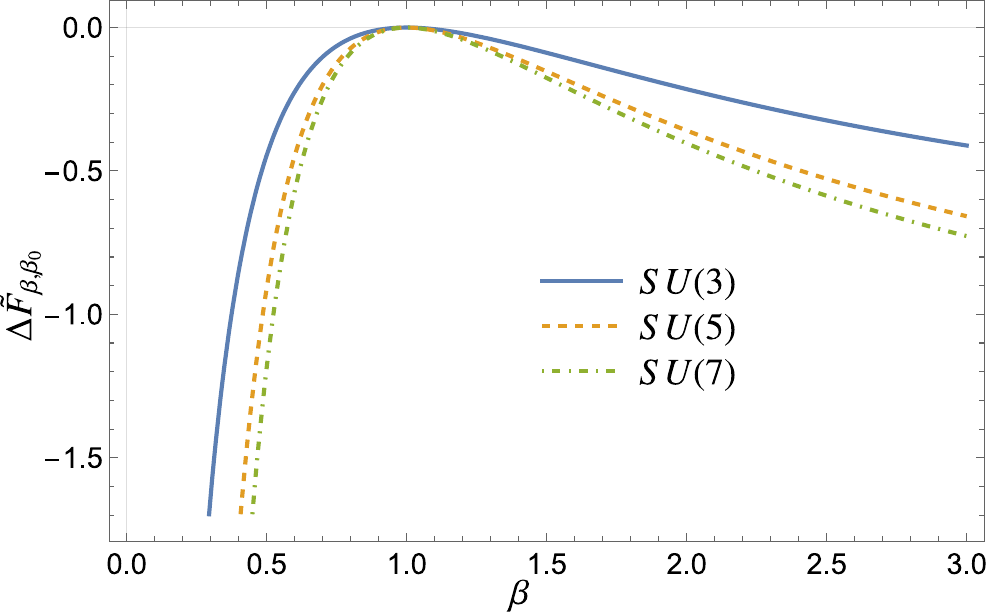}
    \caption{Plot of the free energy change \eqref{equ:FreeEnergyChange} as a function of the inverse bath temperature $\beta$ for an ensemble of 5 particles and initial temperature $\beta_0=1$ for $\mathrm{SU}(d)$ ensembles with $d = 3,5,7$.}
    \label{fig:FreeEnergyD}
\end{figure}

As we have seen in the previous sections, the general tools developed in this paper can be used to calculate thermodynamical quantities for non-interacting spin-$s$ ensembles and ensembles with arbitrary $\mathrm{SU}(d)$-symmetry. 
Beyond these applications, our framework also makes it possible to analyse the anatomy of the steady states of such ensembles on the level of individual irreps.

\subsection{Free energy and dimensions}

To illustrate this approach, which can provide further insights into the role of higher-order symmetries, we first consider the change in the reduced non-equilibrium free energy as defined in Eq.~\eqref{equ:FreeEnergyChange}.  
For a quantitative comparison between the different symmetry groups, we choose the coefficients $a_i$ in the single-particle Hamiltonian $h= \sum_i a_i d_i$ to reproduce the ladder spectrum $\varepsilon_i = -(d+1)/2 + i$. 
With this choice we first plot the non-equilibrium free energy change as function of the inverse bath temperature $\beta$ in Fig.~\ref{fig:FreeEnergyD}. 
We see that, for all bath temperatures, $\abs{\Delta \tilde{F}}$ increases with the order $d$ of the applied symmetry group. 

\begin{figure}[h!]
    \includegraphics[width=.45\textwidth]{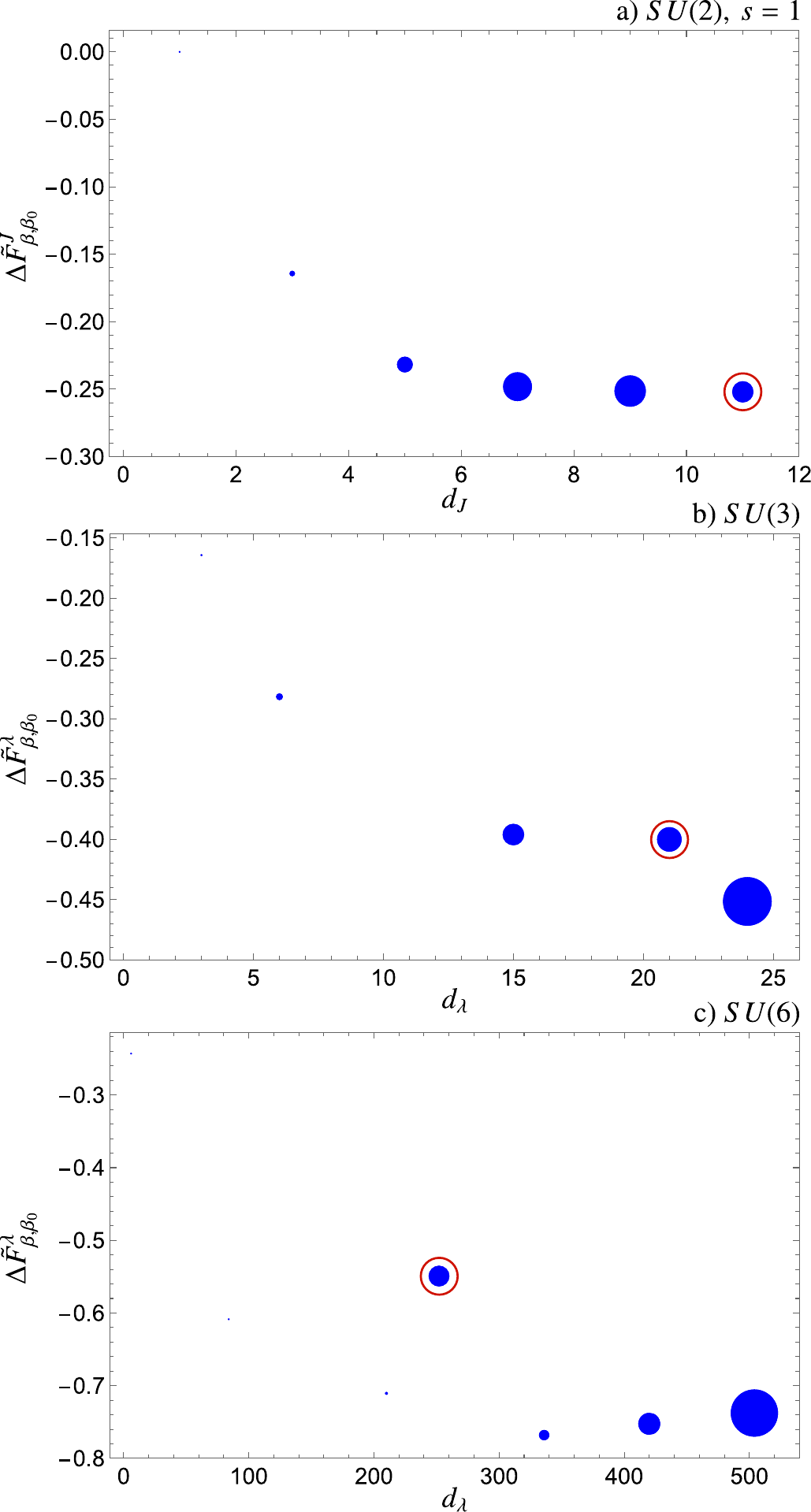}
    \caption{Contributions to the free energy change $ \Delta \tilde{F}^\lambda_{\beta,\beta_0}$ for each irrep $\lambda$, such that $\Delta \tilde{F}_{\beta,\beta_0} = \sum_\lambda p^\lambda_{\beta_0} \Delta \tilde{F}^\lambda_{\beta,\beta_0}$, as functions of the irrep dimension $d_\lambda$.
    The radius of each dot is scaled in proportion with the corresponding occupation probability $p^\lambda_{\beta_0}$. 
    In the first plot, which corresponds to an $\mathrm{SU}(2)$-ensemble with spin $s=1$, the maximal $J$ irrep is circled. 
    In the middle and bottom plots, which correspond to $\mathrm{SU}(3)$- and $\mathrm{SU}(6)$-ensembles, the circle indicates the symmetric irrep. 
    The system parameters are $n=5,\, \beta_0 = 1,\, \beta=3$. 
    \label{fig:free_energy_terms}}
\end{figure}

To understand this effect better, we now focus on the role of the individual irreps. 
In Fig.~\ref{fig:free_energy_terms}, we decompose $\Delta \tilde{F}_{\beta,\beta_0} = \sum_\lambda p^\lambda_{\beta_0} \Delta \tilde{F}^\lambda_{\beta,\beta_0}$ into contributions from the individual irreps $\lambda$; the size of each dot in these plots indicates the relative probability of occupying a particular irrep.
For spin ensembles, the maximal angular momentum irrep $J = ns$, which has the largest dimension $d_\text{max}=2ns + 1$, is the predominantly occupied irrep in the limit $\beta_0\rightarrow\infty$. 
For $\mathrm{SU}(d)$, the symmetric irrep, which does not generally have the largest dimension, is most occupied for large $\beta_0$. 
Thus, for spin ensembles, the dominant contribution to the free energy change can only shift to a lower-dimensional irrep as $\beta_0$ decreases to a moderate value -- see Fig.~\ref{fig:free_energy_terms}. 
By contrast, for $\mathrm{SU}(d)$-ensembles with $d>2$, one observes a shift towards a higher-dimensional irrep.

\subsection{Degeneracies}

An explanation for the link between thermodynamical properties and irrep dimension can be given in terms of energy level degeneracies.
Here, one finds a notable difference between $\mathrm{SU}(2)$ and higher-order groups.
$\mathrm{SU}(2)$ irreps are special in having non-degenerate energy levels -- that is, for each $J$, the energy basis $\ket{J,M}$ varying over $M$ is non-degenerate.
For $\mathrm{SU}(3)$ and above, we observe two sources of degeneracy.
The first is that different sets of occupation numbers $n_i$ of the single-particle energy levels $\varepsilon_i$ may have the same total energy.
This can only happen when the $\varepsilon_i$ are rationally dependent.
For example, with the three-level ladder where $(\varepsilon_i)_i = (-1,0,1)$, the occupation numbers $(0,2,1)$ and $(1,0,2)$ represent three-particle configurations with the same energy of $+1$.
The ability of the higher-order group dynamics to mix between such configurations results from the greater number of dissipation channels, compared with a single channel for $\mathrm{SU}(2)$. 

The second type of degeneracy occurs \emph{within} a given set of occupation numbers, and can be seen by considering Young tableaux.
Referring to the construction described in Sec.~\ref{sec:young_diagrams} of a basis for the $\mathrm{SU}(d)$ irreps, we consider the example $d=3,\, n=3$ and the irrep for $\lambda = (2,1,0)$.
Two of the possible Young tableaux associated with this irrep are
\begin{align}
    \begin{ytableau}
        1 & 2 \\ 
        3
    \end{ytableau}\;, \quad 
    \begin{ytableau}
        1 & 3 \\ 
        2
    \end{ytableau}\;.
\end{align} 
These Young tableaux describe two linearly independent, though non-orthogonal, states in the subspace $\mc{H}^\lambda$.
By orthogonalising, one obtains a pair of basis states with the same occupation numbers $(n_i)_i = (1,1,1)$ but which are still distinguishable according to the permutation-invariant system dynamics.
It is not hard to see that such situations can only occur with $n>2$ and $d>2$.
Moreover, only symmetries other than fully symmetric and anti-symmetric can support this degeneracy -- with only a single row or column in the Young diagram, the filling order is fixed.
This is why (as in Fig.~\ref{fig:free_energy_terms}) the symmetric irrep is not the one with highest dimension, other than for $\mathrm{SU}(2)$.

This second type of degeneracy is intrinsically non-classical in the sense that states of classical indistinguishable particles are labelled only by their occupation numbers. 
Thus, for more than two particles and dynamics generating a higher-order symmetry group, the effective state space contains degrees of freedom that do not exist in ensembles of classical indistinguishable particles.
Notably, the same mechanism underlies the quantum modification recently found in the well-known Gibbs paradox~\cite{Yadin2021Mixing}.

Are these additional degeneracies useful?
Whereas the ground and fully excited states are non-degenerate (see App.~\ref{app:ground_state}), the degeneracies tend to increase in the middle of the spectrum.
The work output of the Otto cycle in Sec.~\ref{sec:engine} is dictated by the heat capacity $C = V(E)/T^2$, where $V(E)$ is the variance of the energy~\cite{Latune2020-ex}.
This can be seen when $\kappa \bc$ is close to $\bh$ by relating the energy difference $E_{\bh, \beta_0} - E_{\kappa \bc, \beta_0}$ to its derivative with respect to bath temperature, using $C = \partial E / \partial T$.
Thus the work output is proportional to the energy variance.
For high bath temperatures $T_\text{c}$ and $T_\text{h}$, the populations are roughly uniform over the energy spectrum, and so the advantage gained by the $\mathrm{SU}(2)$ system is due to its suppression of degeneracies in the middle of the spectrum, thus raising the energy variance.
In contrast, for low bath temperatures, most of the population is in the ground state, so raising the degeneracy of higher energy states will increase the variance.
Therefore distinguishable particles will perform better in this case.
For an intermediate temperature range, one expects $\mathrm{SU}(3)$ to perform best, since it has degeneracies at higher energies, though not as many as for distinguishable particles.
These observations thus qualitatively explain the behaviour in Fig.~\ref{fig:work_otto}.

It remains to be explored whether the additional non-classical degrees of freedom can be further exploited to gain an additional advantage in some setting.
For example, one could engineer a Hamiltonian that breaks these degeneracies in order to extract work from them.
Unlike superradiance, which has been argued to exist in classical models~\cite{Gross1982Superradiance,Nefedkin2016Superradiance}, this effect would have no classical analogue.

\section{Perspectives} \label{sec:perspectives}

In this paper, we have combined methods from representation theory to build a comprehensive theoretical framework for the thermodynamical description of non-interacting quantum many-body systems with permutation-invariant observables. 
As our main application of this formalism, we have investigated the structure and properties of steady states that emerge when such systems are weakly coupled to a thermal environment. 
We have further shown that permutation invariance induces collective effects that can in principle be used to enhance the performance of quantum thermal machines. 
At every step of our analysis, we have demonstrated that altering the constituents of a permutation-invariant ensemble from spin systems, which are covered by the conventional Clebsch-Gordan theory, to more general multi-level systems, for which we have developed a systematic description in this paper, can lead to qualitative changes in the thermodynamical properties of these ensembles. 

Two possible extensions of our approach appear to be promising perspectives for future research. 
First, it would be interesting to include oscillating driving fields. 
For fast driving, this extension can be achieved by replacing steady-state master equations with thermodynamically consistent Floquet-Lindblad equations \cite{Alicki2006Internal,Kosloff2013}, which have been used earlier to study permutation-invariant spin systems \cite{Niedenzu_2018}. 
In the slow-driving regime, one can instead employ adiabatic master equations, whose generators are obtained by replacing time-independent system parameters with external control protocols \cite{Alicki1979Quantum,Brandner2016,Dann2018Time,Albash2012Quantum}. 
In this way, it would in particular be possible to explore the role of permutation-invariance in the context of thermodynamic geometry, a topic that is currently attracting much interest in quantum thermodynamics \cite{Miller_2019,Scandi_2019,Brandner_2020,Pancotti_2020,Abiuso_2020,Miller_2020,Miller_2021,
Eglinton_2022}.

From a mathematical perspective, progress in these directions would require an extension of our theorem on the structure of steady states arising from autonomous permutation-invariant master equations to periodic limit cycles. 
Since the generators of Floquet-Lindblad equations become time independent in a rotating basis, this generalisation should be straightforward in the fast-driving regime. 
For the adiabatic limit, a counterpart of Spohn's theorem \cite{Spohn1977} for steady states that provides uniqueness conditions for periodic limit cycles is available \cite{Menczel_2019}. 
This result will however still have to be generalised to permutation-invariant many-body systems. 
Similarly, it is important to generalise the setting to regimes beyond weak coupling, involving techniques such as stochastic Liouville equations~\cite{Vadimov2021Validity} and the thermal leads approach~\cite{Lacerda2022Quantum}.
Permutation symmetry can in principle be included in such settings, so the main theoretical challenge is to find simple models under which the steady state is the same partially thermalised state analysed here.

A second key problem is to systematically investigate the role of interactions.  
Most of our results, in particular our theorem on steady states, depend only on permutation invariance and should therefore be applicable also to interacting systems. 
However, since interactions typically lead to a dense energy spectrum, they render the rotating wave approximation that underpins the conventional weak-coupling master equation invalid. 
It would thus be necessary to either identify a relevant class of interacting systems that still feature a spectrum with well-separated Bohr frequencies, or to employ new types of recently derived quantum master equations \cite{Nathan_2020,Mozgunov_2020,Davidovic2020Completely}, whose thermodynamical consistency is however not yet settled.
One class of interacting Hamiltonians which may be most tractable is those with a product state energy basis -- such as pairwise $d_i d_j$ interactions -- since the eigenvectors of $H^\lambda$ can be taken as the Schur basis.

From an information-theoretical viewpoint, one extension could involve studying the full work statistics of the Otto cycle rather than just the mean -- due to the conservation law relating to the symmetry type, each block has its own associated probability and work output, so one is forced to consider the work as a fluctuating random variable.
The thermodynamical quantities considered here would then need to be replaced by single-shot versions~\cite{Dahlsten2011Inadequacy}.
For example: what is the minimal work output that is guaranteed per cycle with some small probability of failure?
Another line of questioning is what our results have to say about the thermodynamical processing of ensembles beyond the ``independent and identically distributed" (i.i.d.) regime, where one typically considers many independent copies of a state subject to arbitrary operations.
This setting has been used to collapse the many quantum ``second laws"~\cite{Brandao2015Second} down to just the standard non-equilibrium free energy~\cite{Brandao2013}.
Perhaps collective operations as defined here may give rise to different laws.

\acknowledgments
B.Y.\ acknowledges helpful discussions with Stefan Nimmrichter, Julia Boeyens, Camille Latune and Sai Vinjanampathy.
K.B.\ acknowledges support from the University of Nottingham through a Nottingham Research Fellowship. This work was supported by
the Medical Research Council [grant number MR/S034714/1]; and the Engineering and Physical Sciences Research Council
[grant numbers EP/V031201/1, EP/T517902/1].
This project has received funding from the European Union's Framework Programme for Research and Innovation
Horizon 2020 (2014-2020) under the Marie Sk\l odowska-Curie Grant Agreement No. 945422.\\

\section*{Data availability}
All plotted data were generated from the equations discussed in the text. No further data were created by the research presented in this paper.

\bibliography{main}

\newpage
\onecolumngrid
\appendix 

\section{Steady states of non-interacting ensembles with $\mathrm{SU}(d)$-symmetry} \label{app:steady_state}

The main result of this appendix is \textbf{Theorem 1}. 
This theorem specifies the structure of the steady states of a class of collective quantum master equations, which is more general than the one discussed in the main text.  

We consider a permutation-invariant ensemble of $n$ $d$-level particles embedded in a large, not necessarily thermal, bath. 
The bare ensemble is described by a permutation-invariant Hamiltonian $H$, which may in principle also contain interactions between particles. 
We further assume that the system-bath interaction is described by the Hamiltonian $H_I = \hbar\Delta\sum_\mu E_\mu\otimes B_\mu$, where the $E_\mu$ are the non-diagonal generators of the tensor-product representation of $\mathrm{SU}(d)$ on the ensemble Hilbert space and the Hermitian operators $B_\mu=B_{-\mu}$ correspond to bath observables. 
The operators $E_\mu$ can now be decomposed into Bohr-frequency components with respect to the ensemble Hamiltonian $H$, that is, $E_\mu = \sum_\omega E_\mu(\omega)$ with $[H,E_\mu(\omega)]= \hbar \omega E_\mu(\omega)$. 
Provided that the conventional weak-coupling, Born-Markov and rotating wave approximations are applicable, the time evolution of the state $\rho$ of the ensemble can then be described in terms of the master equation $\partial_t \rho_t = \mathcal{L}(\rho_t)$ with the Lindblad generator $\mathcal{L}$ being defined in Eq.~\eqref{appA:Lindblad}. 

Before we can move on to the proof of our main theorem, we need a lemma that characterises those operators that commute with all $E_\mu$.
Note that no information is needed about the dimensions of multiplicity spaces $\mathcal{K}^\lambda$, so that the results below apply not just when Schur-Weyl duality is valid, but also to spin systems upon replacing $\lambda$ by $J$.

\begin{lemma} \label{lem:commutators}
    Let $\{\mu\}$ be a set of simple roots of $\mathrm{SU}(d)$.
    Then, for an operator $X$ it holds that
    \begin{equation}
        [X, E_\mu] = [X, E_{-\mu}] = 0 \; \forall \mu \qq{if and only if}  X = \bigoplus_\lambda X^\lambda \otimes \id_{\mc{H}^\lambda},
    \end{equation}
    where $X^\lambda$ is an arbitrary operator on the multiplicity space $\mc{K}^\lambda$.
    \begin{proof}
        The ``if" part is obvious.
        For the ``only if" part, we make use of the fact that the simple roots plus their negatives generate the whole root system (see Ref.~\cite{FultonHarris}, Eq.~21.20).
        In other words, for any root $\nu$, there exists a sequence of simple roots or their negatives, $\mu^{(1)},\mu^{(2)},\dots, \mu^{(k)}$ (allowing for repetitions) such that
        \begin{align}
            E_\nu \propto [E_{\mu^{(k)}}, [\dots, [E_{\mu^{(2)}}, E_{\mu^{(1)}}] \dots] ].
        \end{align}
        If $[X, E_{\mu^{(i)}}] = 0 \; \forall i$, we therefore see that $[X, E_\nu] = 0$, as $E_\nu$ is constructed from products of operators commuting with $X$.

        For the remaining diagonal generators, we make use of the commutation relation $[E_\mu, E_{-\mu}] = \sum_i M_{\mu i} D_i$.
        There is a choice of normalisation such that $M_{\mu i} = v_\mu^i$~\cite[Chapter VI.3]{Zee2016}.
        Then, for any simple root $\mu$ in the set,
        \begin{align} \label{eqn:diag_commutators}
            \sum_i v_\mu^i [X, D_i] & = [X, [E_\mu, E_{-\mu}]] \nonumber \\
                & = [E_\mu, [E_{-\mu},X]] + [E_{-\mu},[X,E_\mu]] \nonumber \\
                & = 0,
        \end{align}
        having used the Jacobi identity in the second line.
        Since the $d-1$ simple roots are a linearly independent set, \eqref{eqn:diag_commutators} can only hold for all $\mu$ if all commutators $[X, D_i]$ vanish.
        Therefore we have found that $X$ must commute with the representation of $\mathrm{SU}(d)$.

        Finally, we decompose the unitary representation as $U = \bigoplus_\lambda \id_{\mc{K}^\lambda} \otimes U^\lambda$.
        Considering any block of $X^{\lambda,\lambda'}$ with respect to the irrep structure, commutation implies $(\id_{\mc{K}^\lambda} \otimes U^\lambda) X^{\lambda,\lambda'} (\id_{\mc{K}^{\lambda'}} \otimes {U^{\lambda'}}^\dagger) = X^{\lambda,\lambda'}$.
        Schur's lemma then tells us that the off-diagonal blocks vanish, while the diagonal ones are of the form $X^{\lambda, \lambda} = X^\lambda \otimes \id_{\mc{H}^\lambda}$.
    \end{proof}
\end{lemma}

\begin{theorem} \label{thm:steady_state}
    Consider the Lindblad generator
    \begin{equation}\label{appA:Lindblad}
        \mc{L}(\rho) = -\frac{i}{\hbar}[H,\rho] + \sum_{\mu,\nu} \Gamma^{\mu \nu}_\omega \left[ E_\mu(\omega) \rho E_\nu(\omega) ^\dagger - \frac{1}{2} \{ E_\nu(\omega)^\dagger E_\mu(\omega), \rho \} \right],
    \end{equation}
    where the $E_\mu$ are the collective representations of the non-diagonal generators of the Lie algebra $su(d)$ in the Cartan basis, $H$ is any permutation-invariant Hamiltonian and the coefficients $\Gamma^{\mu \nu}_\omega$ form a Hermitian matrix for each Bohr frequency $\omega$.
    Further assume that there exists a set of simple roots $R$ such that, for each $\omega$, the coefficients $\Gamma^{\mu \nu}_\omega$ form a matrix that has full rank over those roots $\mu \in R$ for which the frequency components $E_\mu(\omega)$ do not vanish. 
    Then the steady states of the generator $\mathcal{L}$ are all of the form
    \begin{equation} \label{eqn:ss_form}
        \rho_\infty = \bigoplus_\lambda p^\lambda \sigma^\lambda \otimes \rho^\lambda,
    \end{equation}
    where the probabilities $p^\lambda$ and the operators $\sigma^\lambda$ are arbitrary, while the $\rho^\lambda$ are unique on each irrep $\lambda$.
    \begin{proof}

        We adapt the original proof by Spohn that guarantees a single steady state \cite{Spohn1977}, more closely following the presentation of this proof in Ref.~\cite{Rivas2014Quantum}.
        First, we show that a set of $\rho^\lambda$ exists such that all states of the form \eqref{eqn:ss_form} are steady states.
        Due to the permutation symmetry of the dynamics, any such state must evolve in time as $\rho_t = \bigoplus_\lambda p^\lambda \sigma^\lambda \otimes \rho^\lambda_t$.
        Since the irreps $\mc{H}^\lambda$ are finite-dimensional, each contains at least one steady state $\rho^\lambda$.\\

        Next, we still have to prove that this family covers all the steady states.
        Initially, we relabel the set $\{E_\mu(\omega)\}_{\mu,\omega}$ as $\{E_i\}_i$ such that $i$ corresponds to a pair $(\mu_i, \omega_i)$.
        Next we put the generator into diagonal form by transforming to $F_i = \sum_j u_{ji} E_j$, where the $u_{ji}$ form a unitary matrix.
        Then we have $E_i = \sum_j u_{ij}^* F_j$, so
        \begin{align}
            \mc{L}(\rho) & = -\frac{i}{\hbar} [H,\rho] + \sum_{i,j} \delta_{\omega_i,\omega_j} \Gamma_{\omega_i}^{\mu_i \mu_j} \left[ E_i \rho E_j^\dagger - \frac{1}{2} \{ E_j^\dagger E_i, \rho \} \right] \nonumber \\
            & = -\frac{i}{\hbar} [H,\rho] + \sum_{k,l} A_{k l} \left[ F_k \rho F_l^\dagger - \frac{1}{2} \{ F_j^\dagger F_i, \rho \} \right],
        \end{align}
        with $A_{k l} = \sum_{i,j} u_{ik}^* \delta_{\omega_i, \omega_j} \Gamma_{\omega_i}^{\mu_i \mu_j} u_{jl}$.
        Choosing $u$ to diagonalise the Hermitian matrix $A$, we obtain
        \begin{align}
            \mc{L}(\rho) = -\frac{i}{\hbar} [H,\rho] + \sum_k a_k \left[ F_k \rho F_k^\dagger - \frac{1}{2} \{F_k^\dagger F_k, \rho \} \right].
        \end{align}
        We would like to guarantee that $a_k > 0$.
        To this end, we note that $A$ is unitarily equivalent to the matrix 
        \begin{equation}
            \Gamma = \bigoplus_\omega \Gamma_\omega := \bigoplus_\omega \sum_{\substack{\mu, \nu: \\ E_\mu(\omega), E_\nu(\omega) \neq 0}} \Gamma_\omega^{\mu \nu} \dyad{\omega, \mu}{\omega, \nu}.
        \end{equation}
        Note that we use Dirac notation here for convenience although the vectors $\ket{\omega,\mu}$ do not represent quantum states.
        The aim of this construction is to include only those terms for which $E_\mu$ has a nonzero frequency component $E_\mu(\omega)$; otherwise, the full-rank condition would not generally be possible to satisfy.
        Thus we see that the positivity of $A$ is guaranteed by assuming that all $\Gamma_\omega$ have full rank.

        We then make a second transformation by expanding $F_i = \sum_{i=1}^p f_{i k} G_k$ in some operator basis such that the $G_k$ are Hermitian and the $f_{i k}$ are complex.
        The $G_k$ are chosen to form a basis of the subspace $V = \mathrm{span}\{F_i\}_i = \mathrm{span}\{E_\mu\}_\mu$ with some dimension $p$.
        This substitution results in
        \begin{align}
            \mc{L}(\rho) & := -\frac{i}{\hbar} [H,\rho] + \sum_{k,l=1}^p B_{k l} \left[ G_k \rho G_l - \frac{1}{2}\{ G_l G_k, \rho\} \right],
        \end{align}
        where $B_{k l} = \sum_i a_i f_{i k} f_{i k}^*$.
        Evidently $B = \sum_i a_i \vb{f}_i \vb{f}_i^\dagger > 0$, which means we can find some $b>0$ smaller than the smallest eigenvalue of $B$ such that $B - b\id_p > 0$.
        Using this result, we divide $\mc{L}$ into two parts:
        \begin{align}
            \mc{L} & := \mc{L}_1 + \mc{L}_2, \\
            \mc{L}_2(\rho) & := b \sum_{k=1}^p G_k \rho G_k - \frac{1}{2} \{G_k^2, \rho\},
        \end{align}
        such that both parts are valid Lindblad generators.

        We first analyse the spectral properties of $\mc{L}_2$.
        For any Hermitian operator $X$, we calculate the super-operator expectation value
        \begin{align} \label{eqn:l2_expval}
            \ev{\mc{L}_2}_X = \tr{X \mc{L}_2(X)} & = b \sum_k \tr{ X G_k X G_k - \frac{1}{2}X G_k^2 X - \frac{1}{2} X^2 G_k^2 } \nonumber \\
                & = \frac{b}{2} \sum_k \tr{ [G_k,X]^2 } \nonumber \\
                & = -\frac{b}{2} \sum_k \|[G_k,X]\|_2^2.
        \end{align}
        This quantity is generally negative and vanishes if and only if $[G_k, X] = 0 \; \forall i=1,\dots,p$.
        This condition is equivalent to $[E_\mu,X]=0 \; \forall \mu$, and by Lemma~\ref{lem:commutators}, also equivalent to $X = \bigoplus_\lambda X^\lambda \otimes \id_{\mc{H}^\lambda}$.
        Let us denote the subspace of such operators by $S = \{\bigoplus_\lambda X^\lambda \otimes \id_{\mc{H}^\lambda} \mid {X^\lambda}^\dagger = X^\lambda \}$; this space has dimension $M = \sum_\lambda m_\lambda (m_\lambda + 1)/2$.
        We have already found a space of steady states with this dimension, so our goal is to show that $\mc{L}$ has precisely $M$ linearly independent zero eigenvectors.

        Next we divide the vector space of operators into two orthogonal parts $S \oplus S'$.
        There is then a block decomposition of $\mc{L}$ on a suitable basis respecting this structure: representing the super-operator as a matrix,
        \begin{align}
            \mc{L} = \begin{pmatrix}
                * & * \\ * & \mc{L}'
            \end{pmatrix},
        \end{align}
        and similarly for $\mc{L}_{1,2}$, where $*$ denotes an unspecified block and $\mc{L}'$ is the projection onto $S'$.
        From \eqref{eqn:l2_expval} we see that $\ev{\mc{L}_2}_X < 0$ for all $X \in S'$, so the eigenvalues of $\mc{L}'_2$ all have strictly negative real part.
        Similarly, for all $X \in S'$,
        \begin{align}
            \ev{\mc{L}'}_X = \ev{\mc{L}'_1}_X + \ev{\mc{L}'_2}_X <0,
        \end{align}
        since $\mc{L}'_1$ must have eigenvalues with non-positive real part in order to be a valid Lindblad generator.
        Therefore $\mc{L}'$ also has eigenvalues with strictly negative real part.

        Finally, to learn about the eigenvalues of $\mc{L}$, we use Theorem 1.4.10 of Ref.~\cite{Horn1985Matrix}.
        This theorem tells us that if an eigenvalue $\lambda$ of some $n \times n$ matrix has geometric multiplicity of at least $k+1$, then every $(n -k) \times (n-k)$ principal submatrix also has $\lambda$ as an eigenvalue.
        Suppose for a contradiction that $\mc{L}$ has more than $M$ zero eigenvectors.
        So we apply this theorem with $\lambda=0,\, k=M$ and $n$ as the dimension of the full operator space.
        This observation implies that $\mc{L}'$ has a zero eigenvector, which contradicts what we have just determined about the eigenvalues of $\mc{L}'$.
        Therefore $\mc{L}$ has at most $M$ zero eigenvectors, and we have already found this many.
    \end{proof}
\end{theorem}

With a thermal environment, one can assume a standard detailed-balance relation on the bath correlation function and thereby obtain the following result guaranteeing thermalisation within each block:

\begin{corollary} \label{cor:steady_state_thermal}
    With the same assumptions as in Theorem~\ref{thm:steady_state}, together with $\Gamma^{\mu \nu}_{-\omega} = e^{\beta \hbar \omega} \Gamma^{\nu \mu}_\omega$, the steady states are all those of the form
    \begin{equation}
        \rho_\infty = \bigoplus_\lambda p^\lambda \sigma^\lambda \otimes \gamma^\lambda,
    \end{equation}
    where $\gamma^\lambda = e^{-\beta H^\lambda}/\mathrm{tr}[e^{-\beta H^\lambda}]$ is the thermal state on each irrep, according to the Hamiltonian $H = \bigoplus_\lambda \id_{\mc{K}^\lambda} \otimes H^\lambda$.
    \begin{proof}
        This result is an immediate consequence of Theorem~\ref{thm:steady_state} followed by applying, to each diagonal block, Spohn's result~\cite{Spohn1977} that the condition $\Gamma^{\mu \nu}_{-\omega} = e^{\beta \hbar \omega} \Gamma^{\nu \mu}_\omega$ requires the steady state $\rho^\lambda$ to be the thermal state $\gamma^\lambda$.
    \end{proof}
\end{corollary}

\section{Symmetric subspace energies}\label{app:steady_state_sym_E}
Here we study the mean energy for the symmetric subspace at high bath temperature to first order in $\beta$, for general $n$ and $d$.
We consider an arbitrary set of single-particle energies $\varepsilon_i,\, i=1,\dots,d$.
A natural basis for the symmetric subspace is given by symmetrising $\bigotimes_{i=1}^d \ket{i}^{\otimes n_i}$ for each configuration $\bm{n} = (n_1,n_2,\dots,n_d),\, \sum_i n_i = n$.
These are energy eigenstates with energy $\varepsilon_{\bm{n}} = \sum_i \varepsilon_i n_i$.

The mean energy when confined to the symmetric subspace is then
\begin{align}
    E^\mathrm{sym}_\beta = \frac{\sum_{\bm{n}} \varepsilon_{\bm{n}} e^{-\beta \varepsilon_{\bm{n}}}}{\sum_{\bm{n}} e^{-\beta \varepsilon_{\bm{n}}}},
\end{align}
which is, to first order at high temperature,
\begin{align}
    E^\mathrm{sym}_\beta & \approx \frac{1}{d_\text{sym}} \sum_{\bm{n}} \varepsilon_{\bm{n}} - \beta \left[ \frac{1}{d_\text{sym}} \sum_{\bm{n}} \varepsilon_{\bm{n}}^2 - \left( \frac{1}{d_\text{sym}} \sum_{\bm{n}} \varepsilon_{\bm{n}} \right)^2 \right] \nonumber \\
    & = \ev{\varepsilon}_n - \beta \llangle\varepsilon^2\rrangle_n,
\end{align}
where $d_\text{sym} = \binom{n+d-1}{n}$, and $\ev{\varepsilon}_n,\, \llangle\varepsilon^2\rrangle_n$ denote the average and variance of energies, respectively, in a uniform mixture for $n$ particles.
We will express these in terms of their values for $n=1$.

The mean energy is
\begin{align}
    \ev{\varepsilon}_n & = \sum_i \ev{n_i}_n  \varepsilon_i \nonumber \\
    & = \ev{n_1}_n \sum_i \varepsilon_i \nonumber \\
    & = \frac{n}{d} \sum_i \varepsilon_i \nonumber \\
    & = n \ev{\varepsilon}_1,
\end{align}
having used the fact that the uniform distribution of energies is symmetric with respect to permutations of the energy levels.
Similarly,
\begin{align}
    \ev{\varepsilon^2}_n & = \sum_{i,j} \ev{n_i n_j}_n \varepsilon_i \varepsilon_j \nonumber \\
    & = \sum_i \ev{n_i^2}_n \varepsilon_i^2 + \sum_{i\neq j} \ev{n_i n_j}_n \varepsilon_i \varepsilon_j \nonumber \\
    & = \ev{n_1^2}_n \sum_i \varepsilon_i^2 + \ev{n_1 n_2}_n \left( \sum_{i,j} \varepsilon_i \varepsilon_j - \sum_i \varepsilon_i^2 \right) \\ \nonumber
    & = \ev{n_1^2}_n d \ev{\varepsilon^2}_1 + \ev{n_1 n_2}_n \left( d^2 \ev{\varepsilon}^2_1 - d \ev{\varepsilon^2}_1 \right).
\end{align}
For simplicity, we can take $\ev{\varepsilon}_1 = 0$ without loss of generality -- then we have
\begin{align} \label{eqn:var_1_and_n}
    \llangle\varepsilon^2\rrangle_n = \ev{\varepsilon^2}_n = \ev{\varepsilon^2}_1 \, d \left( \ev{n_1^2}_n - \ev{n_1 n_2}_n \right).
\end{align}
Given that $\sum_i n_i = n$, we have
\begin{align}
    n^2 & = \sum_{i,j} \ev{n_i n_j}_n \nonumber \\
    & = \sum_i \ev{n_i^2}_n + \sum_{i\neq j} \ev{n_i n_j}_n \nonumber \\
    & = d \ev{n_1^2}_n + d(d-1) \ev{n_1 n_2}_n.
\end{align}
Substituting into Eq.\eqref{eqn:var_1_and_n}, we have
\begin{align} \label{eqn:var_1_and_n_ratio}
    \frac{\llangle\varepsilon^2\rrangle_n}{\llangle\varepsilon^2\rrangle_1} & = \frac{d^2 \ev{n_1^2}_n}{d-1} - \frac{n^2}{d-1}.
\end{align}
The probability distribution over $n_1$ is found by counting the number of ways of distributing the remaining $n-n_1$ particles over the other $d-1$ energy levels:
\begin{align}
    p(n_1) = \frac{\binom{n-n_1+d-2}{d-2}}{\binom{n+d-1}{d-1}},
\end{align}
from which we get
\begin{align}
    \ev{n_1^2}_n = \sum_{n_1=0}^n n_1^2 \, p(n_1) = \frac{n(2n+d-1)}{d(d+1)}.
\end{align}
Eq.~\eqref{eqn:var_1_and_n_ratio} then gives
\begin{align} \label{eqn:var_ratio_result}
    \frac{\llangle\varepsilon^2\rrangle_n}{\llangle\varepsilon^2\rrangle_1} = \frac{n(n+d)}{d+1}.
\end{align}
This is to be compared with the full (infinite temperature) thermal state of $n$ particles, which has energy variance $n \llangle\varepsilon^2\rrangle_1$.\\

We can now ask what set of single-particle energies gives the maximal variance; due to Eq.~\eqref{eqn:var_ratio_result}, we only need to consider the case of a single particle.
In order to make a well-posed question, we constrain the $\varepsilon_i$ such that there is at least one at the minimum energy $\varepsilon_1 = 0$ and at least one at the maximum $\varepsilon_d = d-1$.
(These values are chosen to match the spread of the ladder Hamiltonian -- note that the overall shift can be arbitrary.)
The variance must always satisfy
\begin{align}
    \llangle\varepsilon^2\rrangle_1 \leq \frac{(\varepsilon_d-\varepsilon_1)^2}{4} = \frac{(d-1)^2}{4}.
\end{align}
For even $d$, this can be saturated by taking half the levels at $0$ and half at $d-1$.
For odd $d$, the optimum is with either $\frac{d-1}{2}$ at $0$ and $\frac{d+1}{2}$ at $d-1$, or the opposite.
Hence,
\begin{align}
    \max_{\{\varepsilon_i\},\, \varepsilon_1=0,\, \varepsilon_d=d-1} \llangle\varepsilon^2\rrangle_1 =
    \begin{cases}
    \frac{(d-1)^2}{4}, & d \text{ even},\\
    \frac{(d-1)^2}{4} \left(1-\frac{1}{d^2}\right), & d \text{ odd}.
    \end{cases}
\end{align}
For the evenly spaced ladder, one instead has $\llangle\varepsilon^2\rrangle_1 = \frac{d^2-1}{12}$ -- so the level optimisation roughly gains a factor of three.

\section{Ground state in each subspace}\label{app:ground_state}
Here, we argue that each term $H^\lambda$ of the Hamiltonian has a unique ground state (on the reduced state space).
This is seen by examining the construction of the irreps of $\mathrm{SU}(d)$ described in Sec.~\ref{sec:young_diagrams}.
Given a single-particle basis $\ket{k}$ -- which we take as the single-particle energy eigenbasis -- one can construct a (generally non-orthogonal) basis of the irrep corresponding to a Young diagram $\lambda$ by specifying all ways of filling the boxes of $\lambda$ according to certain rules.
Each box must be filled with a label $k$ in a way that is weakly increasing (i.e., two neighbouring elements can be the same) in each row and strictly increasing in each column.
For example, one allowed tableau with $n=8,\, d=3$ is
\begin{align}
    \begin{ytableau}
    1 & 1 & 2 & 3\\
    2 & 3 & 3 \\
    3
    \end{ytableau}.
\end{align}
Each allowed tableau corresponds to a vector constructed by taking a product state of each $\ket{k}$ per box and applying the Young symmetriser, which symmetrises over each row and then anti-symmetrises over each column.
Each such state is an energy eigenstate with eigenvalue $\sum_{i=1}^n \varepsilon_{k_i}$, where $k_i$ is the label of box $i$ in the tableau.
It is clear that there is only one way to construct a state of minimal energy: fill each column of length $l$ with the labels $1,2,\dots,l$ from top to bottom.
For the above diagram, this gives
\begin{align}
    \begin{ytableau}
        1 & 1 & 1 & 1 \\
        2 & 2 & 2 \\
        3
    \end{ytableau}.
\end{align}
The unique ground state in the irrep $\lambda$ has energy
\begin{align} \label{eqn:lambda_energy}
E_\lambda = \sum_{k=1}^d \lambda_k \varepsilon_k.
\end{align}
Note that this argument depends crucially on the Hamiltonian being non-interacting.

\section{Asymptotic energies and entropies for $\mathrm{SU}(d)$} \label{app:asymptotics}
Here we find the mean energy of a state with $\beta_0=0$ in the limit of large particle number $n\to \infty$.
We use the ladder Hamiltonian with energies
\begin{align}
    \varepsilon_k = -\frac{(d+1)}{2} + k, \quad k=1,\dots,d,
\end{align}
and the limiting probability distribution $\phi_d(\zeta)$ in Eq.~\eqref{equ:kerov_limits} with $\zeta_i = \frac{\lambda_i - n/d}{\sqrt{n}}$.
From Eq.~\eqref{eqn:lambda_energy} together with $\sum_k \zeta_k = 0$ and $\sum_k \varepsilon_k=0$, we have the ground state energy in the block $\lambda$,
\begin{align} \label{eqn:lambda_energy_asymp}
    E_\lambda & = \sum_{k=1}^d \left(\sqrt{n} \zeta_k + \frac{n}{d} \right) \varepsilon_k \nonumber \\
        & = \sqrt{n} \sum_{k=1}^d \zeta_k \left(k - \frac{d+1}{2} \right) \nonumber \\
        & = \sqrt{n} \sum_{k=1}^d k \zeta_k \nonumber \\
        & = \sqrt{n} \left( \sum_{k=1}^{d-1} k \zeta_k + d\zeta_d \right) \nonumber \\
        & = \sqrt{n} \sum_{k=1}^{d-1} (k-d) \zeta_k.
\end{align}
We have chosen to eliminate $\zeta_d$ because of the constraint $\sum_k \zeta_k=0$, leaving only the remaining $(d-1)$ variables independent.
The limiting mean energy is then
\begin{align}
    E_{\infty,0} & = \sqrt{n} \int \dd^{d-1} \zeta  \; \phi_d(\zeta) \sum_{k=1}^{d-1}(k-d)\zeta_k \\
        & =: - \sqrt{n} \mc{E}_d,
\end{align}
where the integral is constrained to $\zeta_k \geq \zeta_{k+1} \; \forall k$.
We find the coefficients
\begin{align}
    \mc{E}_2 = \sqrt{\frac{2}{\pi}}, \quad \mc{E}_3 = \frac{9}{4} \sqrt{\frac{3}{\pi}}.
\end{align}
Higher dimensions are only numerically tractable; we find
\begin{align}
    \mc{E}_2 \approx 0.798,\, \mc{E}_3 \approx 2.20,\, \mc{E}_4 \approx 4.19,\, \mc{E}_5 \approx 6.76,\, \mc{E}_6 \approx 9.91,\, \mc{E}_7 \approx 13.6.
\end{align}
For the entropy $\tilde{S}_{0,0}$, we need to find the average of $\ln d_\lambda$.
Given that $\lambda_i - \lambda_j = \sqrt{n}(\zeta_i - \zeta_j)$, one can see from Eq.~\eqref{equ:dimUlambda} that
\begin{align}
    d_\lambda \propto \prod_{i<j} \left( \sqrt{n}[\zeta_i - \zeta_j] + [i-j] \right)
\end{align}
Since the distribution $\phi_d(\zeta)$ is independent of $n$, the function to be averaged contains $d(d-1)/2$ terms of the order $\ln \sqrt{n}$.
We therefore have, to leading order in $n$,
\begin{align}
    \tilde{S}_{0,0} = \ev{ \ln d_\lambda } \approx \frac{d(d-1)}{4} \ln n.
\end{align}

\section{Asymptotics for $\mathrm{SU}(2)$, spin-1} \label{app:spin_asym}
Here we consider the same limit as above, with an initially high temperature $\beta_0 = 0$ in the limit of large $n$, for $d=3$ and $\mathrm{SU}(2)$-type coupling to the bath.
We will find the limiting distribution $p^J$.
For $n$ particles with general spin $s$, the thermal state $\gamma_0$ on the full Hilbert space is maximally mixed, so one sees that
\begin{equation}
    p^J = \frac{(2J+1) m_J}{(2s+1)^n}.
\end{equation}
We use the recursion relation \eqref{equ:spinrec} for $m_J$ to derive a similar relation for $p^J$.
Writing the probability explicitly as a function of $n$, we have
\begin{align}
    p(n+1, J) = \sum_{\substack{J' :\; |J'-s|\leq J\leq J' +s,\\ J'+s-J \, \in\,  \mathbb{Z}}} \frac{(2J+1)}{(2s+1)(2J'+1)} p(n, J).
\end{align}
From now on assuming $s=1$, we have
\begin{align} \label{eqn:spin1_rec}
    p(n+1, J) = \begin{cases}
        \frac{p(n, 1)}{9} , & J = 0 \\
        \frac{2J+1}{3} \left[ \frac{p(n,J-1)}{2J-1} + \frac{p(n,J)}{2J+1} + \frac{p(n,J+1)}{2J+3} \right], & J = 1,2,\dots .
    \end{cases}
\end{align}
with only integer values of $J$ appearing.
In the limit of large $n$, we make the ansatz that $p(n,J)$ takes the form
\begin{align}
    p(n,J) \approx \frac{1}{\sqrt{n}} f\left( \frac{J}{\sqrt{n}} \right),
\end{align}
where $f$ is some smooth function normalised such that $\int_0^\infty f(x) \, \dd x = 1$.
This guess will be justified by showing that it approximately satisfies the recursion relation \eqref{eqn:spin1_rec}, and we will determine $f$.
For any $J \geq 1$, \eqref{eqn:spin1_rec} reduces to
\begin{align}
    \frac{1}{\sqrt{n+1}} f\left(\frac{J}{\sqrt{n+1}}\right) = \frac{1}{3\sqrt{n}} \left[ f\left( \frac{J}{\sqrt{n}}\right) + \left(\frac{2J+1}{2J-1}\right) f\left( \frac{J-1}{\sqrt{n}}\right) +  \left(\frac{2J+1}{2J+3}\right) f\left( \frac{J+1}{\sqrt{n}}\right) \right].
\end{align}
We write $\delta = 1/\sqrt{n}$, $x = J \delta$, giving
\begin{align}
    \frac{1}{\sqrt{1+\delta^2}} f\left(\frac{x}{\sqrt{1+\delta^2}} \right) = \frac{1}{3} \left[ f(x) + \left(\frac{2x+\delta}{2x-\delta} \right) f(x-\delta) + \left( \frac{2x+\delta}{2x+3\delta}\right) f(x+\delta) \right].
\end{align}
Since $\delta \ll 1$, we expand to lowest non-vanishing order in $\delta$:
\begin{align}
    0 = \frac{\delta^2}{3} \left[ -f''(x) + \left( \frac{2}{x} -\frac{3x}{2}\right)f'(x) - \left(\frac{2}{x^2} + \frac{3}{2}\right) f(x) \right] + \mc{O}(\delta^3).
\end{align}
Hence we obtain the second-order linear differential equation
\begin{align}
    2f''(x) + \left(3x -\frac{4}{x}\right) f'(x) + \left(\frac{4}{x^2} + 3\right) f(x) = 0.
\end{align}
The only normalisable solution is found to be $f(x) \propto x^2 e^{-3x^2/4}$, resulting in
\begin{align} \label{eqn:spin1_prob}
    p(n, J) \approx \frac{3\sqrt{3}}{2\sqrt{\pi}} \frac{J^2}{n^{\frac{3}{2}}} e^{- \frac{3J^2}{4n}}.
\end{align}
Since the ground state energy of the irrep $J$ is $-J$, we find the limiting mean energy with $\beta = \infty$ as
\begin{align}
    E_{\infty, 0} & \approx -  \int_0^\infty \dd J \; J\, p(n, J) \nonumber \\
        & \approx - \sqrt{n} \int_0^\infty \dd x \; x f(x) \nonumber \\
        & = -\frac{4}{\sqrt{3\pi}} \sqrt{n}.
\end{align}
Similarly, the $\beta=0$ reduced entropy is
\begin{align}
    \tilde{S}_{0,0} & \approx \int_0^\infty \dd x \; \ln(2 \sqrt{n} x + 1) f(x) \nonumber \\
        & \approx \frac{1}{2} \ln n.
\end{align}

\end{document}